\documentclass[%
reprint,
amsmath,amssymb,
aps,
pra,
floatfix,
]{revtex4-2}

\usepackage{array}[=2016-10-06]

\usepackage[dvipdfmx]{graphicx}
\usepackage{dcolumn}
\usepackage{bm}

\usepackage{color}

\begin{document}
\setlength\textfloatsep{6pt}


\title{
Generation of coherent-state Schr{\"{o}}dinger cat states in a
two-photon Jaynes-Cummings model implemented with Josephson junctions
}

\author{Hiroo Azuma$^{1}$}
\email{hiroo.azuma@oist.jp}
\author{William J. Munro$^{1}$}
\author{Kae Nemoto$^{1, 2}$}
\affiliation{$^1$Okinawa Institute of Science and Technology Graduate University, Onna-son, Okinawa 904-0495, Japan}
\affiliation{$^2$Global Research Center for Quantum Information Science,
National Institute of Informatics, 2-1-2 Hitotsubashi, Chiyoda-ku, Tokyo 101-8430, Japan}




\date{\today}

\begin{abstract}
In this paper, we investigate a method for generating coherent-state
Schr{\"{o}}dinger cat states using a two-photon Jaynes-Cummings model (JCM)
implemented in a Josephson-junction quantum circuit.
So far, traditionally, two main approaches have been considered for producing coherent-state cat states:
a method based on quantum nondemolition measurement, and a probabilistic
photon-subtraction technique.
Although these two methods have been technically improved year by year, neither has yet reached the stage of practical application.
The generation method based on the two-photon JCM investigated in this paper utilizes a superconducting flux-qubit circuit that has already undergone substantial development, and is therefore expected to present relatively low barriers to practical implementation.
In this paper, we examine a specific implementation of a flux qubit and estimate the coupling constant of the two-photon JCM induced by this implementation.
We also examine three types of noise that may arise in this superconducting Josephson-junction circuit: random magnetic flux noise with $1/f$ fluctuations, energy relaxation noise that causes the $T_{1}$ decay of the flux qubit, and photon-loss noise in the bosonic mode obtained by quantizing the phase variable of the dc SQUID used to read out the flux qubit. We show that the third type of noise, that is, the photon loss, is particularly serious.
In the latter part of this paper, we examine how the probability-amplitude distribution of the Pegg-Barnett phase-operator eigenstates evolves in time for the
$l$-photon JCM ($l=1,2,3,4$) when the bosonic mode is prepared in a coherent state initially.
Moreover, preparing a displaced thermal coherent state initially, then disconnecting it from the heat bath, and evolving the system unitarily,
we study how introducing a temperature parameter into the bosonic mode affects the fidelity and the Wigner function of the cat states obtained in experiments.
\end{abstract}

\maketitle


\section{\label{section-introduction}Introduction}

Implementing quantum-optical circuits that generate non-Gaussian states has
long been considered difficult by many researchers, and work on this topic has continued for many years \cite{Walschaers2021}.
A representative example of non-Gaussian states is the Schr{\"{o}}dinger cat state.
In particular, the state $|\alpha\rangle +e^{i\theta}|-\alpha\rangle$, formed from a coherent state $|\alpha\rangle$ with a large average photon number
$|\alpha|^{2}\gg 1$ and the opposite-phase coherent state $|-\alpha\rangle$ with a specific angle $\theta$, can be regarded as a superposition of macroscopically distinct states.
From the viewpoint of theoretical physics, it is an extremely interesting object.
For example, we can regard this macroscopic quantum superposition as supporting evidence for Everett's many-worlds interpretation \cite{EverettIII1957,DeWitt1970}.

As methods for generating the coherent-state cat state, two main approaches have traditionally been studied.
The first approach uses quantum nondemolition (QND) measurement techniques \cite{Yurke1987,Brune1992,Vitali1997,Gerry2006}.
This method is characterized by relying on the cross-Kerr interaction between two photon modes.
In general, the coupling constant of the cross-Kerr interaction is proportional to the third-order nonlinear susceptibility $\chi^{(3)}$.
However, the value of $\chi^{(3)}$ in real nonlinear media is extremely small.
As a result, the cross-Kerr interaction available in practice is very weak, and it is difficult to experimentally demonstrate QND measurements.
As a method for overcoming this problem, the enhancement of nonlinear optical effects using epsilon-near-zero materials has been actively studied in recent years; however, it has not yet become a key component in the generation of non-Gaussian states \cite{Fomra2024}.

Recently, however, a high-fidelity QND readout protocol for generating cat states has been reported in a theoretical study \cite{Xu2026}.
This method is characterized by the use of a Kerr-nonlinear resonator with an effective longitudinal interaction.
Although this report concerns QND measurement techniques and, strictly speaking, does not directly discuss the generation of coherent-state cat states, it predicts that a fidelity of $99.99$\% can be achieved with a measurement time of $50$ [ns].
Furthermore, $\alpha=5$ is given as an example of the coherent-light amplitude for which the protocol proposed in \cite{Xu2026} operates effectively.

The second approach is photon subtraction \cite{Dakna1997,Wenger2004,Ourjoumtsev2006,Neergaard-Nielsen2006}.
This method consists of the following steps.
First, squeezed light is generated and sent into a beam splitter, where it is divided into two beams.
Next, a single-photon measurement is performed at one of the two output ports of the beam splitter.
When this single-photon detection is used as a heralding signal, the light emerging from the other output port becomes an approximate cat state.
This method is highly effective for generating cat states with a small average photon number
but it is not suitable for producing states with a large average photon number.
For this reason, the states produced by the photon-subtraction technique are called the Schr{\"{o}}dinger's kitten states.
An attempt to generate large-photon-number cat states using the photon-subtraction method was reported in Ref.~\cite{Ourjoumtsev2007}.

As a recent experimental report on the photon-subtraction method, Ref.~\cite{Endo2026} is worth mentioning.
In this work, multi-photon generalized photon subtraction in picosecond optical wave packets was demonstrated.
An effective cat-state amplitude of $\alpha=1.69$ was achieved, and the experiment was performed at a pump repetition rate of $5$ [MHz].
Thus, although the two conventional methods for generating cat states have been steadily improved, their practical implementation is still likely to require considerable time.

As described above, the two conventional methods have greatly contributed to the analyses of non-Gaussian states, but they are not capable of generating cat states with a large average photon number.
There is, however, another method for generating the cat states that did not attract much attention from the research community.
This method produces the coherent-state Schr{\"{o}}dinger cat states using the resonant two-photon Jaynes-Cummings model (JCM) \cite{Dung1994,Nasreen1994}.
The method consists of the following steps.
First, an atom is placed inside an optical cavity, and the atom is prepared in its ground state while the cavity field is set to a coherent state; this serves as the initial state.
The angular frequency of the atom is precisely tuned to twice the angular frequency of the cavity photons.
Next, the combined atom-cavity system is allowed to evolve in time according to the two-photon JCM.
After a certain period, a projective measurement is performed on the atom, distinguishing between the ground and excited states.
This measurement collapses the cavity field into the coherent-state Schr{\"{o}}dinger cat state.

A more detailed summary of the two papers \cite{Dung1994,Nasreen1994} is as follows.
If the coupling constant between the atom and the cavity field in the two-photon JCM is denoted by $g_{2}$, then after a time $t=\pi/(2g_{2})$, the measurement of the atom yields a cat state.
Reference \cite{Dung1994} clearly states this result.
In contrast, Ref.~\cite{Nasreen1994} explicitly describes the cavity-field states at
$t=\pi/(4g_{2})$, $\pi/g_{2}$, and $2\pi/g_{2}$, but does not examine the cavity-field state at $t=\pi/(2g_{2})$.
This omission occurred because the authors of Ref.~\cite{Nasreen1994} traced out the cavity-field degrees of freedom and found that the resulting atomic density matrix becomes maximally mixed at $t=\pi/(2g_{2})$.
They therefore judged that this particular case was not especially important.

Although this method was sufficiently new and original, it did not attract much attention among researchers in 1994.
This was likely because, at that time, no concrete or practical experimental implementation of the two-photon JCM had been found.
However, in recent years, rapid progress in quantum-information-processing devices has led to serious consideration of how to realize the two-photon JCM in practice.
In particular, methods based on a superconducting Josephson-junction circuit for a flux quantum bit (qubit) were recognized as promising for implementing the two-photon JCM \cite{Felicetti2018}.
The magnetic flux qubit in the Josephson circuit is one of the leading candidates for realizing a quantum computer, and it has been actively studied for many years \cite{Nakamura1999,Clarke2008,Arute2019}.

In addition to the above, Ayyash {\it et al}. proposed a method for generating cat states using a superconducting circuit \cite{Ayyash2024}.
They proposed a cat-state generation scheme based on a continuously driven qubit in a superconducting resonator.
In their effective model, the coherent-state amplitude grows as $|\alpha|=g_{1}t/2$ from an initial vacuum state, where $g_{1}$ is a coupling constant of the single-photon JCM.
In the method presented in Refs.~\cite{Dung1994,Nasreen1994}, the coherent-state amplitude $|\alpha|$ is already set in the initial state of the cavity field, and the two methods contrast in this respect.

In this paper, we propose generating coherent-state Schr{\"{o}}dinger cat states using the resonant two-photon JCM implemented in a superconducting Josephson-junction circuit that works as the flux qubit.
With this approach, we can generate cat states whose average photon number is large with a high generation rate.
We estimate the coupling constant $g_{2}$ on the basis of experimentally obtained values.
We also examine three types of noise that may pose obstacles in experiments: random magnetic flux noise with $1/f$ fluctuations, energy relaxation noise that causes the $T_{1}$ decay of the flux qubit, and photon-loss noise in the bosonic mode obtained by quantizing the phase variable of the dc SQUID used to read out the flux qubit.
The first type of noise is treated as a time-dependent coherent error, whereas the second and third types are treated as dissipation described by master equations.
Our numerical calculations lead to the conclusion that the third type of noise, that is, the photon loss, is serious.

In addition, this paper examines the resonant $l$-photon JCM ($l=1,2,3,4$) by decomposing the cavity-field state into superpositions of the eigenvectors of the Pegg-Barnett phase operator, and by studying how the probability-amplitude distribution of these eigenstates evolves over time \cite{Pegg1988,Pegg1989,Barnett1989}.
We find that the peak of the phase-amplitude distribution, which is initially single, splits into two as time progresses, later merges back into one, and then splits again---showing a periodic behavior.
Moreover, for $l=1,3,4$, the wave packet of the probability distribution spreads out over time and loses its sharpness, whereas in the two-photon JCM with $l=2$, the wave packet remains sharp even as time passes.
Finally, introducing a temperature parameter into the $l$-photon JCM with a preparation of an initial thermal coherent state for photons and evolving the system unitarily, we confirm that the fidelity of the approximate cat states obtained in experiments decreases
as the temperature increases.

The phenomenon described above---where the phase-component distribution splits from one peak into two---had already been pointed out in earlier studies of the single-photon JCM \cite{Eiselt1991,Jex1993}.
The results in this paper extend those earlier findings.
In this paper, we examine the $l$-photon JCM ($l=2,3,4$), which is a generalization of the usual single-photon JCM.
One of the characteristic features of these models, the collapse and revival of Rabi oscillations, plays an important role here.
Numerical calculations in this paper show that the period of the phase-distribution oscillation in the $l$-photon JCM coincides with the period of the collapse and revival of Rabi oscillations.
For a comprehensive review of the JCM, see Ref.~\cite{Shore1993}.
The derivation of the collapse-and-revival period for the $l$-photon JCM is given in Ref.~\cite{Azuma2024}.

In this paper, we focus on the fact that cat states can be obtained from the time evolution of the two-photon JCM when the initial cavity field is prepared as a coherent state.
Reference \cite{Gea-Banacloche1991} studied the single-photon JCM with the cavity field initially prepared in a coherent state, and showed that a state resembling a cat state appears at certain times during the evolution.
There was also a study of the two-photon JCM in which the initial cavity-field state was already a cat state, and its time evolution was analyzed \cite{Fakhri2021}.
However, the problem setting in Ref.~\cite{Fakhri2021} is different from that of this paper.

This paper is organized as follows.
In Sec.~\ref{section-eigenvalues-eigenstates-mJCM},
we provide a review about eigenvalues and eigenstates of the Hamiltonian of $l$-photon JCM for $l=1,2,3,4$.
In Sec.~\ref{section-generation-cat-state},
we provide a review of the method for generating the cat states with two-photon JCM.
In Sec.~\ref{section-experiment},
we explain experimental implementation of the two-photon JCM with the flux qubit in the Josephson-junction circuit.
In Sec.~\ref{section-noise},
we investigate effects of three typical noises induced in the quantum circuit.
In Sec.~\ref{section-time-evolution-phase},
we numerically estimate time evolution of the phase of the cavity field in the multiphoton JCM.
In Sec.~\ref{section-temperature},
we compute the detection probability and fidelity of the approximately generated cat state under thermal noise.
In Sec.~\ref{section-conclusion},
we provide conclusion.

\section{\label{section-eigenvalues-eigenstates-mJCM}Eigenvalues and eigenstates of the Hamiltonian of the multiphoton JCM}

In this section, we consider the Hamiltonian of the
$l$-photon JCM for $l=1,2,...$,
which generalizes the standard JCM.
Now, we derive its eigenvalues and eigenvectors.
The Hamiltonian of the multiphoton JCM is given by
\begin{equation}
\hat{H}
=
\frac{\hbar\omega_{0}}{2}\hat{\sigma}_{z}
+
\hbar\omega\hat{a}^{\dagger}\hat{a}
+
\hbar g_{l}[\hat{\sigma}_{+}\hat{a}^{l}+\hat{\sigma}_{-}(\hat{a}^{\dagger})^{l}],
\end{equation}
where $\omega_{0}$ denotes the angular frequency corresponding to the difference in energies of the excited and ground states of the atom, $\omega$ represents the angular frequency of the cavity field, $g_{l}$ denotes the coupling constant,
and $l=1, 2, 3, ...$.
Here, note that $g_{l}$ has the dimension of the angular frequency.
Moreover, the annihilation and creation operators of the cavity field, $\hat{a}$ and $\hat{a}^{\dagger}$, satisfy the commutation relation ${[}\hat{a},\hat{a}^{\dagger}{]}=1$, and the $z$-component Pauli matrix and the lowering and raising operators of the atom are written in the forms,
\begin{equation}
\hat{\sigma}_{z}
=
\left(
\begin{array}{cc}
1 & 0\\
0 & -1\\
\end{array}
\right),
\ 
\hat{\sigma}_{-}
=
\left(
\begin{array}{cc}
0 & 0\\
1 & 0\\
\end{array}
\right),
\ 
\hat{\sigma}_{+}
=
\left(
\begin{array}{cc}
0 & 1\\
0 & 0\\
\end{array}
\right),
\end{equation}
respectively,
where the ground and excited states of the atom are given by
$|g\rangle_{\mbox{\scriptsize A}}=(0,1)^{\mbox{\scriptsize T}}$
and
$|e\rangle_{\mbox{\scriptsize A}}=(1,0)^{\mbox{\scriptsize T}}$, respectively.
The subscript A indicates that the state belongs to the atom.
Introducing the interaction picture, we divide the Hamiltonian $\hat{H}$ as
\begin{eqnarray}
\hat{H}
&=&
\hat{C}_{1}+\hat{C}_{2}, \nonumber \\
\hat{C}_{1}
&=&
\hbar\omega[(l/2)\hat{\sigma}_{z}+\hat{a}^{\dagger}\hat{a}], \nonumber \\
\hat{C}_{2}
&=&
\hbar(-\Delta/2)\hat{\sigma}_{z}+\hbar g_{l}[\hat{\sigma}_{+}\hat{a}^{l}+\hat{\sigma}_{-}(\hat{a}^{\dagger})^{l}], \nonumber \\
\Delta
&=&
-\omega_{0}+l\omega.
\label{interaction-picture-Hamiltonians-0}
\end{eqnarray}
Because $[\hat{C}_{1},\hat{C}_{2}]=0$,
we can describe a wave function of the system with the interaction picture as
\begin{equation}
|\Psi_{\mbox{\scriptsize I}}(t)\rangle=\hat{U}(t)|\Psi_{\mbox{\scriptsize I}}(0)\rangle,
\label{interaction-picture-1}
\end{equation}
where
\begin{eqnarray}
\hat{U}(t)
&=&
\exp(-i\hat{C}_{2}t/\hbar)
=
\left(
\begin{array}{cc}
\hat{u}_{00} & \hat{u}_{01}\\
\hat{u}_{10} & \hat{u}_{11}\\
\end{array}
\right),
\label{unitary-time-evolution-operator-1}
\end{eqnarray}
\begin{eqnarray}
\hat{u}_{00}
&=&
\cos(\sqrt{\hat{D}}t)
+
i\frac{\Delta}{2}
\frac{\sin(\sqrt{\hat{D}}t)}{\sqrt{\hat{D}}}, \nonumber \\
\hat{u}_{01}
&=&
-ig_{l}
\frac{\sin(\sqrt{\hat{D}}t)}{\sqrt{\hat{D}}}\hat{a}^{l}, \nonumber \\
\hat{u}_{10}
&=&
-ig_{l}
\frac{\sin(\sqrt{\hat{D}'}t)}{\sqrt{\hat{D}'}}(\hat{a}^{\dagger})^{l}, \nonumber \\
\hat{u}_{11}
&=&
\cos(\sqrt{\hat{D}'}t)
-
i\frac{\Delta}{2}
\frac{\sin(\sqrt{\hat{D}'}t)}{\sqrt{\hat{D}'}},
\label{unitary-time-evolution-operator-2}
\end{eqnarray}
\begin{eqnarray}
\hat{D}
&=&
(\Delta/2)^{2}+g_{l}^{2}\hat{a}^{l}(\hat{a}^{\dagger})^{l}, \nonumber \\
\hat{D}'
&=&
(\Delta/2)^{2}+g_{l}^{2}(\hat{a}^{\dagger})^{l}\hat{a}^{l}.
\label{unitary-time-evolution-operator-3}
\end{eqnarray}

Eigenvalues and eigenvectors of the Hamiltonian $\hat{C}_{2}$ of the $l$-photon JCM are given by
\begin{eqnarray}
\hat{C}_{2}|\varphi_{n}\rangle
&=&
E_{n}|\varphi_{n}\rangle,
\quad
E_{n}=\hbar\Delta/2,
\quad
|\varphi_{n}\rangle
=
|g\rangle_{\mbox{\scriptsize A}}
|n\rangle_{\mbox{\scriptsize P}} \nonumber \\
&&
\quad
\mbox{for $n=0, 1, ..., l-1$,}
\end{eqnarray}
\begin{eqnarray}
\hat{C}_{2}|\varphi_{n}^{(\pm)}\rangle
&=&
E_{n}^{(\pm)}|\varphi_{n}^{(\pm)}\rangle, \nonumber \\
E_{n}^{(\pm)}
&=&
\pm
\hbar
[(\Delta/2)^{2}+n(n-1)\cdots(n-l+1)g_{l}^{2}]^{1/2}, \nonumber \\
|\varphi_{n}^{(\pm)}\rangle
&=&
c_{1,n}^{(\pm)}
|g\rangle_{\mbox{\scriptsize A}}
|n\rangle_{\mbox{\scriptsize P}}
+
c_{2,n}^{(\pm)}
|e\rangle_{\mbox{\scriptsize A}}
|n-l\rangle_{\mbox{\scriptsize P}} \nonumber \\
&&
\quad
\mbox{for $n=l, l+1, ...$,}
\label{eigenvalue-eigenstate-0}
\end{eqnarray}
\begin{eqnarray}
c_{1,n}^{(\pm)}
&=&
\frac{\Delta_{n}^{(\pm)}}
{
g_{l}
\sqrt{4n(n-1)\cdots(n-l+1)+|\Delta_{n}^{(\pm)}|^{2}/g_{l}^{2}}
}, \nonumber \\
c_{2,n}^{(\pm)}
&=&
\frac{2}
{
\sqrt{4+|\Delta_{n}^{(\pm)}|^{2}/[n(n-1)\cdots(n-l+1)g_{l}^{2}]}
},
\end{eqnarray}
\begin{equation}
\Delta_{n}^{(\pm)}
=
\Delta
\pm
\sqrt{4n(n-1)\cdots(n-l+1)g_{l}^{2}+\Delta^{2}},
\end{equation}
where
the subscript P denotes the cavity photons and
$\{|n\rangle_{\mbox{\scriptsize P}}\}$ represents their Fock states.

\section{\label{section-generation-cat-state}Generation of the cat states using the resonant two-photon
JCM}

In this section, we give a brief review of how to generate coherent-state cat states using the resonant two-photon JCM
\cite{Dung1994,Nasreen1994}.
Assuming $l=2$ and $\Delta=0$, the eigenenergies and eigenstates of the Hamiltonian take the following form,
\begin{equation}
E_{n}^{(\pm)}
\simeq
\pm \hbar g_{2}(n-\frac{1}{2}-\frac{1}{8n}+...)
\quad
\mbox{for $n\gg 1$},
\end{equation}
\begin{equation}
c_{1,n}^{(\pm)}=\pm\frac{1}{\sqrt{2}},
\quad
c_{2,n}^{(\pm)}=\frac{1}{\sqrt{2}}.
\end{equation}
Thus, the eigenvectors $|\varphi_{n}^{(\pm)}\rangle$ exhibit the following time evolution,
\begin{eqnarray}
|\varphi_{n}^{(\pm)}(t)\rangle
&=&
\exp(-iE_{n}^{(\pm)}t/\hbar)
|\varphi_{n}^{(\pm)}\rangle \nonumber \\
&\simeq&
\exp(\pm ig_{2}t/2)\exp(\mp ig_{2}nt)|\varphi_{n}^{(\pm)}\rangle \nonumber \\
&&
\quad
\mbox{for $n\gg 1$}.
\end{eqnarray}
Here, we assume that the initial state of the cavity system is prepared as
$|\Psi(0)\rangle_{\mbox{\scriptsize AP}}=|g\rangle_{\mbox{\scriptsize A}}
|\alpha\rangle_{\mbox{\scriptsize P}}$
where $|\alpha\rangle_{\mbox{\scriptsize P}}$ represents the coherent state.
We can expand this initial state approximately with $\{|\varphi_{n}^{(\pm)}\rangle\}$
in the form,
\begin{eqnarray}
&&
|\Psi(0)\rangle_{\mbox{\scriptsize AP}} \nonumber \\
&=&
\exp(-|\alpha|^{2}/2)\sum_{n=0}^{\infty}\frac{\alpha^{n}}{\sqrt{n!}}
|g\rangle_{\mbox{\scriptsize A}}
|n\rangle_{\mbox{\scriptsize P}} \nonumber \\
&\simeq&
\frac{1}{\sqrt{2}}
\exp(-|\alpha|^{2}/2)\sum_{n=2}^{\infty}\frac{\alpha^{n}}{\sqrt{n!}}
(
|\varphi_{n}^{(+)}\rangle
-
|\varphi_{n}^{(-)}\rangle
),
\end{eqnarray}
for $|\alpha|\gg 1$.
Thus, the time evolution of the above state is approximately given by
\begin{eqnarray}
&&
|\Psi(t)\rangle_{\mbox{\scriptsize AP}} \nonumber \\
&\simeq&
\frac{1}{\sqrt{2}}
\exp(-|\alpha|^{2}/2)\sum_{n=2}^{\infty}\frac{\alpha^{n}}{\sqrt{n!}}
(
e^{ig_{2}t/2}
e^{-ig_{2}nt}|\varphi_{n}^{(+)}\rangle \nonumber \\
&&
-
e^{-ig_{2}t/2}
e^{ig_{2}nt}|\varphi_{n}^{(-)}\rangle
) \nonumber \\
&\simeq&
\frac{1}{2}
e^{ig_{2}t/2}
|g\rangle_{\mbox{\scriptsize A}}
|e^{-ig_{2}t}\alpha\rangle_{\mbox{\scriptsize P}} \nonumber \\
&&
+
\frac{1}{2}
e^{ig_{2}t/2}
|e\rangle_{\mbox{\scriptsize A}}
\exp(-|\alpha|^{2}/2)
\sum_{n=0}^{\infty}
\frac{(e^{-ig_{2}t}\alpha)^{2}}{\sqrt{(n+1)(n+2)}} \nonumber \\
&&
\times
\frac{(e^{-ig_{2}t}\alpha)^{n}}{\sqrt{n!}}
|n\rangle_{\mbox{\scriptsize P}} \nonumber \\
&&
+
\frac{1}{2}
e^{-ig_{2}t/2}
|g\rangle_{\mbox{\scriptsize A}}
|e^{ig_{2}t}\alpha\rangle_{\mbox{\scriptsize P}} \nonumber \\
&&
-
\frac{1}{2}
e^{-ig_{2}t/2}
|e\rangle_{\mbox{\scriptsize A}}
\exp(-|\alpha|^{2}/2)
\sum_{n=0}^{\infty}
\frac{(e^{ig_{2}t}\alpha)^{2}}{\sqrt{(n+1)(n+2)}} \nonumber \\
&&
\times
\frac{(e^{ig_{2}t}\alpha)^{n}}{\sqrt{n!}}
|n\rangle_{\mbox{\scriptsize P}}.
\label{time-evolution-1}
\end{eqnarray}
Here, since $|\alpha|\gg 1$, the terms around $n\simeq|\alpha|^{2}$ give the dominant contribution in the above summation $\sum_{n=0}^{\infty}$.
In this case, we obtain
\begin{equation}
\frac{(e^{\mp ig_{2}t}\alpha)^{2}}{\sqrt{(n+1)(n+2)}}
\simeq
\exp(\mp 2ig_{2}t).
\end{equation}
Therefore, the time evolution of the system can be written as follows:
\begin{eqnarray}
|\Psi(t)\rangle_{\mbox{\scriptsize AP}}
&\simeq&
\frac{1}{2}
|g\rangle_{\mbox{\scriptsize A}}
(
e^{ig_{2}t/2}|e^{-ig_{2}t}\alpha\rangle_{\mbox{\scriptsize P}}
+
e^{-ig_{2}t/2}|e^{ig_{2}t}\alpha\rangle_{\mbox{\scriptsize P}}
) \nonumber \\
&&
+
\frac{1}{2}
|e\rangle_{\mbox{\scriptsize A}}
[
e^{-i(3/2)g_{2}t}|e^{-ig_{2}t}\alpha\rangle_{\mbox{\scriptsize P}} \nonumber \\
&&
-
e^{i(3/2)g_{2}t}|e^{ig_{2}t}\alpha\rangle_{\mbox{\scriptsize P}}
].
\label{time-evolution-2}
\end{eqnarray}

Here, by setting $t=\pi/(2g_{2})$, the time-evolved state approximately becomes the following:
\begin{eqnarray}
|\Psi(\pi/(2g_{2}))\rangle_{\mbox{\scriptsize AP}}
&\simeq&
\frac{1}{2}
e^{i\pi/4}
|g\rangle_{\mbox{\scriptsize A}}
(
|e^{-i\pi/2}\alpha\rangle_{\mbox{\scriptsize P}}
-
i
|e^{i\pi/2}\alpha\rangle_{\mbox{\scriptsize P}}
) \nonumber \\
&&
+
\frac{1}{2}
e^{-3i\pi/4}
|e\rangle_{\mbox{\scriptsize A}}
(
|e^{-i\pi/2}\alpha\rangle_{\mbox{\scriptsize P}} \nonumber \\
&&
+
i
|e^{i\pi/2}\alpha\rangle_{\mbox{\scriptsize P}}
).
\label{2-mode-JCM-cat-state-obtained-1}
\end{eqnarray}
When we perform a projective measurement of the atomic state in the basis
$\{|g\rangle_{\mbox{\scriptsize A}},|e\rangle_{\mbox{\scriptsize A}}\}$,
the following occurs.
If the outcome is $|g\rangle_{\mbox{\scriptsize A}}$, the photon state collapses to
$|e^{-i\pi/2}\alpha\rangle_{\mbox{\scriptsize P}}
-i
|e^{i\pi/2}\alpha\rangle_{\mbox{\scriptsize P}}$.
If the outcome is $|e\rangle_{\mbox{\scriptsize A}}$, the photon state collapses to
$|e^{-i\pi/2}\alpha\rangle_{\mbox{\scriptsize P}}
+i
|e^{i\pi/2}\alpha\rangle_{\mbox{\scriptsize P}}$.
This corresponds to obtaining coherent-state cat states.
In the context of a microwave resonator, the operation corresponding to a $\pi/2$ phase rotation of the coherent state is realized not by an optical wave plate but by a controlled phase-space rotation, produced by free resonator evolution, a frequency shift, or an applied control pulse.
Under such a $\pi/2$ rotation, the photon states transform as
\begin{equation}
|e^{-i\pi/2}\alpha\rangle_{\mbox{\scriptsize P}}
\mp i
|e^{i\pi/2}\alpha\rangle_{\mbox{\scriptsize P}}
\to
|\alpha\rangle_{\mbox{\scriptsize P}}
\mp i
|-\alpha\rangle_{\mbox{\scriptsize P}}.
\end{equation}

Here, let us emphasize the following point.
For any value of $l$,
in the resonant case with $\Delta=0$, and when the photon number is large ($n\gg 1$),
Eq.~(\ref{eigenvalue-eigenstate-0}) shows that
$E_{n}^{(\pm)}\sim\hbar g_{l}n^{l/2}$.
For $l=2$, the spectrum is approximately linear in $n$,
so the coherent state mainly rotates in phase rather than spreading.
For $l\neq 2$, spectral curvature causes phase dispersion.

In this section, we assume that the interaction is switched off
instantaneously at the half-revival time
$t=\pi/(2g_{2})$,
followed immediately by an ideal projective measurement
of the qubit in its energy eigenbasis, $|g\rangle_{\mbox{\scriptsize A}}$ and $|e\rangle_{\mbox{\scriptsize A}}$.
The measurement is assumed to introduce no additional
disturbance to the oscillator beyond the conditioning
associated with the qubit measurement outcome.
The conditional fidelities reported in the remainder of this paper are defined
under these idealized assumptions.

In this section, the discussion was developed under the assumption that $|\alpha|\gg 1$.
The dependence of the fidelity of the generated cat state on $\alpha$ is examined in Fig.~\ref{figure11}.
According to these numerical results, the fidelity is sufficiently close to unity for $|\alpha|>7.0$.
Specific examples of generation of the cat states with $\alpha=5$ and $\alpha=1.69$ were presented in \cite{Xu2026} and \cite{Endo2026}, respectively.
In contrast, the method discussed in this paper becomes effective for $|\alpha|>7.0$, exhibiting a markedly different characteristic.

\section{\label{section-experiment}Experimental feasibility of generation of the cat states with the Josephson-junction circuit}

In this section, we examine the experimental feasibility of the method shown in Sec.~\ref{section-generation-cat-state} with the Josephson-junction quantum circuit
with which we can construct the magnetic flux qubit.
We begin by considering how to implement the two-photon JCM.
Next, we estimate specific values of the physical parameters on the basis of experimental results reported in previous studies.

\begin{figure}
\begin{center}
\includegraphics[width=0.6\linewidth]{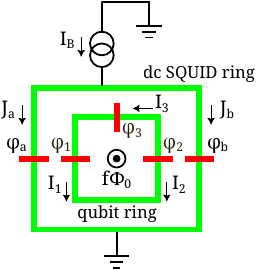}
\end{center}
\caption{
This figure shows the quantum circuit used to construct a superconducting magnetic flux qubit and to read out the state of the qubit.
The red rectangles represent Josephson junctions.
The thick green lines represent leads in the superconducting state.
The quantum circuit consists of two rings, an inner ring and an outer ring.
The inner ring constitutes the flux qubit and contains three Josephson junctions.
The outer ring is a dc SQUID used to read out the state of the flux qubit and contains two Josephson junctions.
An external magnetic flux $f\Phi_{0}$ threads the loop of the flux-qubit ring.
A source of the bias current $I_{\mbox{\scriptsize B}}$ is connected to the dc SQUID.
The bias current $I_{\mbox{\scriptsize B}}$ is used to apply a constant phase externally to the dc SQUID.
}
\label{figure01}
\end{figure}

Figure~\ref{figure01} shows the quantum circuit used to construct a superconducting flux qubit and to read out its state.
The inner ring constituting the superconducting flux qubit in Fig.~\ref{figure01} contains three Josephson junctions.
The idea of constructing a flux qubit with three Josephson junctions was first proposed in Ref.~\cite{Orlando1999}.
The reason for using three Josephson junctions is to form a double-well potential.
In Fig.~\ref{figure01}, let $\varphi_{i}$ denote the phase difference across junction $i$ for $i=1, 2, 3$.
We assume that an external magnetic flux $f\Phi_{0}$ threads the inner circuit in Fig.~\ref{figure01}.
Here, $\Phi_{0}=h/(2e)$ denotes the Dirac magnetic flux quantum.

Owing to the single-valuedness of the superconducting order parameter, the magnetic flux threading the inner ring in Fig.~\ref{figure01} is restricted to integer multiples of $\Phi_{0}$.
Consequently, the phases across the Josephson junctions satisfy the following quantization condition:
\begin{equation}
\varphi_{1}-\varphi_{2}+\varphi_{3}+2\pi f=2\pi n
\quad
(\mbox{$n$: integer}).
\label{quantization-condition-flux-qubit-1}
\end{equation}
The Josephson current flowing through junction $i$ is then given by $I_{i}=I_{0i}\sin\varphi_{i}$,
where $I_{0i}$ is the critical current of junction $i$.
Therefore, the potential energy of the ring is given by
\begin{equation}
U=\sum_{i=1}^{3}E_{\mbox{\scriptsize J}i}(1-\cos\varphi_{i}),
\end{equation}
where $E_{\mbox{\scriptsize J}i}$ is the Josephson coupling energy of junction $i$ and is given by
$E_{\mbox{\scriptsize J}i}=I_{0i}\Phi_{0}/(2\pi)$.
The critical current $I_{0i}$ is proportional to the cross-sectional area of the Josephson junction.
Therefore, the values of $I_{0i}$ and $E_{\mbox{\scriptsize J}i}$ can be freely adjusted by varying the cross-sectional area.
Suppose that we set
$E_{\mbox{\scriptsize J}1}=E_{\mbox{\scriptsize J}2}=E_{\mbox{\scriptsize J}}$
and
$E_{\mbox{\scriptsize J}3}=\alpha E_{\mbox{\scriptsize J}}$.
We then obtain
\begin{equation}
\frac{U}{E_{\mbox{\scriptsize J}}}
=
2+\alpha-\cos\varphi_{1}-\cos\varphi_{2}-\alpha\cos(2\pi f+\varphi_{1}-\varphi_{2}).
\end{equation}

\begin{figure}
\begin{center}
\includegraphics[width=1.0\linewidth]{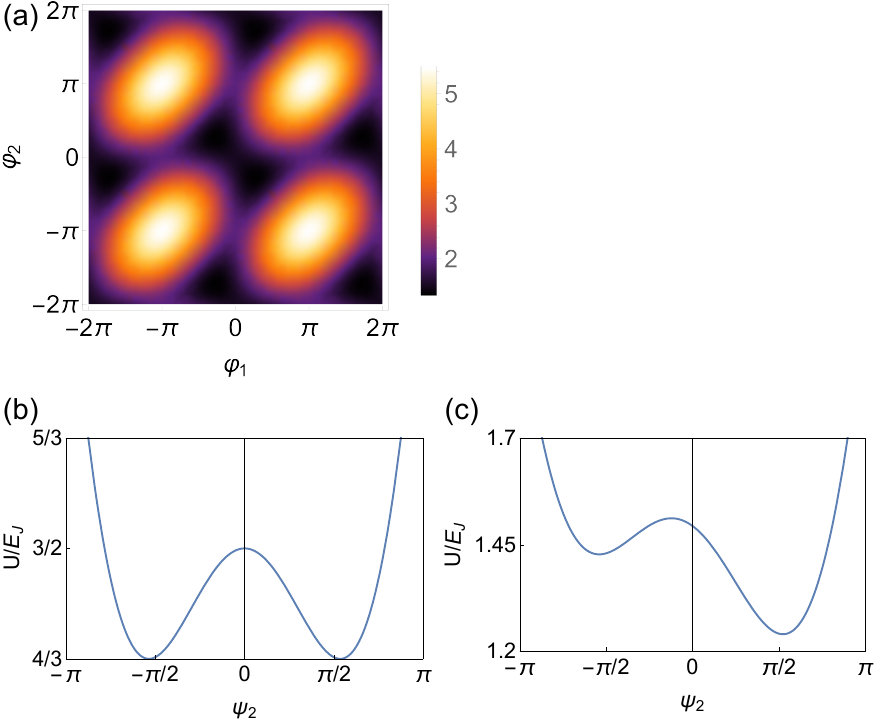}
\end{center}
\caption{
(a)
Plot of $U/E_{\mbox{\scriptsize J}}$ as a function of
$\varphi_{1}$ and $\varphi_{2}$.
A two-dimensional periodic structure is observed.
(b)
Plot of $U/E_{\mbox{\scriptsize J}}$ as a function of
$\psi_{2}=\varphi_{2}-\varphi_{1}$ for
$\alpha=0.75$, $f=1/2$, and $\psi_{1}=\varphi_{1}+\varphi_{2}=0$.
A symmetric double-well structure is observed.
The two minima are located at
$\psi_{2}=\pm 1.68$,
and both have the value $U/E_{\mbox{\scriptsize J}}=4/3$.
The maximum is located at $\psi_{2}=0$, where $U/E_{\mbox{\scriptsize J}}=3/2$.
Therefore, the height of the energy barrier separating the two wells is $E_{\mbox{\scriptsize J}}/6$.
(c)
Plot of $U/E_{\mbox{\scriptsize J}}$ as a function of
$\psi_{2}=\varphi_{2}-\varphi_{1}$ for
$\alpha=0.75$, $f=0.48$, and $\psi_{1}=\varphi_{1}+\varphi_{2}=0$.
An asymmetric double-well structure is observed.
The two minima are located at
$\psi_{2}=-1.70$, $U/E_{\mbox{\scriptsize J}}=1.43$,
and
$\psi_{2}=1.65$, $U/E_{\mbox{\scriptsize J}}=1.24$.
The maximum is located at
$\psi_{2}=-0.381$, $U/E_{\mbox{\scriptsize J}}=1.51$.
}
\label{figure02}
\end{figure}

Figure~\ref{figure02}(a) shows a plot of $U/E_{\mbox{\scriptsize J}}$ as a function of
$\varphi_{1}$ and $\varphi_{2}$ for $\alpha=0.75$ and $f=1/2$.
A two-dimensional periodic structure is observed.
Here, we transform $\varphi_{1}$ and $\varphi_{2}$ into the following two variables:
\begin{equation}
\psi_{1}=\varphi_{1}+\varphi_{2},
\quad
\psi_{2}=\varphi_{2}-\varphi_{1}.
\end{equation}
Figure~\ref{figure02}(b) shows a plot of $U/E_{\mbox{\scriptsize J}}$ as a function of $\psi_{2}$ with $\psi_{1}=0$.
A double-well structure can be observed.
The bottoms of the wells are located at $\psi_{2}=\pm 1.68$, where $U/E_{\mbox{\scriptsize J}}=4/3$.
The barrier separating the two wells is located at $\psi_{2}=0$, where $U/E_{\mbox{\scriptsize J}}=3/2$.
Therefore, the height of the barrier between the two wells is $E_{\mbox{\scriptsize J}}/6$.

Let the Josephson coupling energy and the Cooper-pair charging energy of a qubit junction be $E_{\mbox{\scriptsize J}}=I_{0}\Phi_{0}/(2\pi)$ and $E_{\mbox{\scriptsize C}}=2e^{2}/C$, respectively,
where $I_{0}$ is the critical current and $C$ is the junction capacitance.
For a single Josephson junction at zero bias current,
the plasma angular frequency is given by
$\omega_{\mbox{\scriptsize p}}=\sqrt{2E_{\mbox{\scriptsize J}}E_{\mbox{\scriptsize C}}}/\hbar=\sqrt{2eI_{0}/(C\hbar)}$.
This frequency characterizes small phase oscillations near a minimum of the junction potential.
The corresponding frequencies within each well of the three-junction circuit depend on the effective capacitances and the local curvatures of the circuit potential.
These interwell oscillation frequencies must be distinguished from the transition frequency $\omega_{0}$ of the flux qubit.

The two wells correspond to persistent currents flowing in opposite directions around the qubit loop.
Let $|L\rangle$ and $|R\rangle$ denote the low-energy states localized in the two wells.
Tunneling through the barrier couples these states even at the symmetry point $f=1/2$.
At this point, the ground and first excited states are approximately the symmetric and antisymmetric combinations of $|L\rangle$ and $|R\rangle$,
that is,
\begin{equation}
|0\rangle
=
(1/\sqrt{2})(|L\rangle+|R\rangle),
\quad
|1\rangle
=
(1/\sqrt{2})(|L\rangle-|R\rangle).
\label{LR-01-transformation-1}
\end{equation}
Their energy separation is the tunnel splitting,
denoted by $\Delta_{\mbox{\scriptsize t}}$.
Near $f=1/2$, the qubit Hamiltonian in the localized-state basis can be written as
\begin{equation}
\hat{H}_{\mbox{\scriptsize q}}
=
-\frac{1}{2}
(\epsilon\hat{\sigma}_{z}+\Delta_{\mbox{\scriptsize t}}\hat{\sigma}_{x}),
\end{equation}
where
$\hat{\sigma}_{z}=|L\rangle\langle L|-|R\rangle\langle R|$,
$\hat{\sigma}_{x}=|L\rangle\langle R|+|R\rangle\langle L|$,
and $\epsilon$ is the energy bias between the localized states.
The diagonal energies of the localized states are given by $E_{L}=-\epsilon/2$ and $E_{R}=\epsilon/2$, so that $\epsilon=E_{R}-E_{L}$.
Let the deviation of the external magnetic flux from the symmetry point be
\begin{equation}
\delta\Phi
=
\Phi_{\mbox{\scriptsize ext}}-\frac{\Phi_{0}}{2}
=
\Phi_{0}(f-\frac{1}{2}).
\end{equation}
We define the direction of the persistent current through the dependence of the energy on the external magnetic flux,
$I_{\mbox{\scriptsize p}}=-\partial E/\partial \Phi_{\mbox{\scriptsize ext}}$,
and let the states $|L\rangle$ and $|R\rangle$ carry persistent currents $+I_{\mbox{\scriptsize p}}$ and $-I_{\mbox{\scriptsize p}}$, respectively.
Near the symmetry point, their energies can be approximated to the first order as
$E_{L}\simeq E_{*}-I_{\mbox{\scriptsize p}}\delta\Phi$,
$E_{R}\simeq E_{*}+I_{\mbox{\scriptsize p}}\delta\Phi$,
where $E_{*}$ is their common diagonal energy at the symmetry point.
Therefore,
\begin{equation}
\epsilon
=
E_{R}-E_{L}
\simeq
2I_{\mbox{\scriptsize p}}\delta\Phi,
\end{equation}
which gives
\begin{equation}
\epsilon
\simeq
2I_{\mbox{\scriptsize p}}\Phi_{0}(f-\frac{1}{2}).
\end{equation}
The qubit transition frequency is
$\omega_{0}=\sqrt{\epsilon^{2}+\Delta_{\mbox{\scriptsize t}}^{2}}/\hbar$.
In particular, at $f=1/2$, we have $\epsilon=0$ and $\hbar\omega_{0}=\Delta_{\mbox{\scriptsize t}}$.

As noted above, depending on whether $\psi_{2}=\varphi_{2}-\varphi_{1}$ is positive or negative, counterclockwise or clockwise currents arise, corresponding to the two states
$|L\rangle$ and $|R\rangle$.
When $f=1/2$, the potential energies at the bottoms of the two wells are equal.
However, by shifting the value of $f$ away from $1/2$, a difference can be introduced between the potential energies of the two wells, thereby inducing tunneling between $|L\rangle\leftrightarrow|R\rangle$.
Figure~\ref{figure02}(c) shows a plot of $U/E_{\mbox{\scriptsize J}}$ as a function of $\psi_{2}=\varphi_{2}-\varphi_{1}$ for $\alpha=0.75$, $f=0.48$, and $\psi_{1}=\varphi_{1}+\varphi_{2}=0$.
An asymmetric double-well structure can be observed.
This makes it possible to induce transitions between the states of the flux qubit.

The state of the flux qubit can be detected by measuring the current $I$ flowing through the ring.
More specifically,
when we set $f=1/2$ and $\psi_{1}=0$,
$I$ can be written as
\begin{equation}
I
=
-2I_{0}\sin\frac{\psi_{2}}{2}
-\alpha I_{0}\sin\psi_{2}.
\end{equation}
When $\psi_{2}$ changes from $-1.68$ to $1.68$ where both values are the bottoms of the double well, the current changes as $I\to -I$.
The magnetic flux induced by the qubit current $I$ is proportional to $LI$,
where $L$ is the inductance of the ring constituting the flux qubit.
Therefore, the magnetic flux induced in the flux-qubit ring is reversed depending on whether the state is
$|L\rangle$ or $|R\rangle$.
Consequently, by detecting this small change in the magnetic flux with a dc SQUID, one can distinguish between
$|L\rangle$ and $|R\rangle$ \cite{Felicetti2018,Kittel1996,Feynman2006}.

Next, we consider the quantization of the dc SQUID used to detect the small current flowing through the flux-qubit ring.
As mentioned previously, the inner ring in Fig.~\ref{figure01} is the flux qubit, whereas the outer ring is the dc SQUID.
Let $\hat{\varphi}_{a}$ and $\hat{\varphi}_{b}$ shown in Fig.~\ref{figure01} denote the phase operators across Josephson junctions $a$ and $b$, respectively.
Furthermore, let $\varphi_{\mbox{\scriptsize DC}}$ denote the externally applied constant phase induced by the bias current $I_{\mbox{\scriptsize B}}$.
Let $\hat{\Phi}_{\mbox{\scriptsize tot}}$ denote the total quantized magnetic flux threading the SQUID loop.
In addition, let $\hat{\varphi}_{\mbox{\scriptsize SQ}}$ denote the quantum fluctuation operator of the phase of the dc SQUID resonant mode.
The single-valuedness of the order parameter around the dc SQUID ring leads to the following relations:
\begin{equation}
\hat{\varphi}_{\mbox{\scriptsize SQ}}
=
\hat{\varphi}_{a}
+
\hat{\varphi}_{b},
\quad
2\pi
\left(\frac{\hat{\Phi}_{\mbox{\scriptsize tot}}}{\Phi_{0}}
+
m
\right)
=
\hat{\varphi}_{a}
-
\hat{\varphi}_{b},
\end{equation}
where $m$ is an integer.
Because the term $2\pi m$ vanishes when it is included in the argument of trigonometric functions, we omit it hereafter.
Thus, $\hat{\varphi}_{a}$ and $\hat{\varphi}_{b}$ can be written as
\begin{equation}
\hat{\varphi}_{a}
=
\frac{1}{2}\hat{\varphi}_{\mbox{\scriptsize SQ}}
+
\pi\frac{\hat{\Phi}_{\mbox{\scriptsize tot}}}{\Phi_{0}},
\quad
\hat{\varphi}_{b}
=
\frac{1}{2}\hat{\varphi}_{\mbox{\scriptsize SQ}}
-
\pi\frac{\hat{\Phi}_{\mbox{\scriptsize tot}}}{\Phi_{0}}.
\end{equation}

Let $\hat{J}_{a}$ and $\hat{J}_{b}$ denote the currents flowing through Josephson junctions $a$ and $b$, respectively.
Then, $\hat{J}_{a}$, $\hat{J}_{b}$, and the total current $\hat{J}_{\mbox{\scriptsize tot}}$ flowing through the dc SQUID are given by
\begin{eqnarray}
\hat{J}_{a}
&=&
I_{0}\sin
\left(
\hat{\varphi}_{a}+\frac{1}{2}\varphi_{\mbox{\scriptsize DC}}
\right)
,
\nonumber \\
\hat{J}_{b}
&=&
I_{0}\sin
\left(
\hat{\varphi}_{b}+\frac{1}{2}\varphi_{\mbox{\scriptsize DC}}
\right)
,
\end{eqnarray}
\begin{eqnarray}
\hat{J}_{\mbox{\scriptsize tot}}
&=&
\hat{J}_{a}+\hat{J}_{b} \nonumber \\
&=&
2I_{0}\sin
\left(
\frac{1}{2}\hat{\varphi}_{\mbox{\scriptsize tot}}
\right)
\cos
\left(
\pi
\frac{\hat{\Phi}_{\mbox{\scriptsize tot}}}{\Phi_{0}}
\right),
\end{eqnarray}
where we have defined
$\hat{\varphi}_{\mbox{\scriptsize tot}}
=
\hat{\varphi}_{\mbox{\scriptsize SQ}}+\varphi_{\mbox{\scriptsize DC}}$.
Here, let
$\hat{\Phi}_{q}$ denote
a magnetic flux generated by the effective magnetic dipole moment of the flux qubit,
and let $\Phi_{\mbox{\scriptsize DC}}$ denote
an externally applied static flux.
We then write
$\hat{\Phi}_{\mbox{\scriptsize tot}}
=
\hat{\Phi}_{q}
+
\Phi_{\mbox{\scriptsize DC}}$.

Furthermore, let $M$ denote the qubit-SQUID mutual inductance and $\hat{I}$ the current operator of the qubit.
Then, $\hat{\Phi}_{q}=M\hat{I}$.
Let $I_{\mbox{\scriptsize p}}$ denote the persistent current and define
$\hat{\sigma}_{z}=|L\rangle\langle L|-|R\rangle\langle R|$.
The current operator is then given by
$\hat{I}=I_{\mbox{\scriptsize p}}\hat{\sigma}_{z}$.
The energy $\hat{U}_{\mbox{\scriptsize SQ}}$ of the dc SQUID is given by
\begin{eqnarray}
\hat{U}_{\mbox{\scriptsize SQ}}
&=&
-2E_{\mbox{\scriptsize J}}
\cos
\left(
\pi
\frac{\hat{\Phi}_{\mbox{\scriptsize tot}}}{\Phi_{0}}
\right)
\cos
\left(
\frac{1}{2}\hat{\varphi}_{\mbox{\scriptsize tot}}
\right) \nonumber \\
&=&
-2E_{\mbox{\scriptsize J}}
\left[
\cos
\left(
\pi
\frac{\Phi_{\mbox{\scriptsize DC}}}{\Phi_{0}}
\right)
\cos
\left(
\pi
\frac{\hat{\Phi}_{q}}{\Phi_{0}}
\right)
\right. \nonumber \\
&&
\left.
-
\sin
\left(
\pi
\frac{\Phi_{\mbox{\scriptsize DC}}}{\Phi_{0}}
\right)
\sin
\left(
\pi
\frac{\hat{\Phi}_{q}}{\Phi_{0}}
\right)
\right] \nonumber \\
&&
\times
\left[
\cos
\left(
\frac{\hat{\varphi}_{\mbox{\scriptsize SQ}}}{2}
\right)
\cos
\left(
\frac{\varphi_{\mbox{\scriptsize DC}}}{2}
\right)
\right.
\nonumber \\
&&
\left.
-
\sin
\left(
\frac{\hat{\varphi}_{\mbox{\scriptsize SQ}}}{2}
\right)
\sin
\left(
\frac{\varphi_{\mbox{\scriptsize DC}}}{2}
\right)
\right].
\label{U-SQ-1}
\end{eqnarray}
We now make the following approximations:
\begin{eqnarray}
&&
|\langle\hat{\Phi}_{q}\rangle|/\Phi_{0}\ll 1,
\quad
\cos(\pi\hat{\Phi}_{q}/\Phi_{0})\simeq 1, \nonumber \\
&&
\sin(\pi\hat{\Phi}_{q}/\Phi_{0})\simeq \pi\hat{\Phi}_{q}/\Phi_{0},
\end{eqnarray}
\begin{eqnarray}
&&
|\langle\hat{\varphi}_{\mbox{\scriptsize SQ}}\rangle|\ll\pi,
\quad
\cos(\hat{\varphi}_{\mbox{\scriptsize SQ}}/2)
\simeq
1-(1/8)\hat{\varphi}_{\mbox{\scriptsize SQ}}^{2}, \nonumber \\
&&
\sin(\hat{\varphi}_{\mbox{\scriptsize SQ}}/2)
\simeq
\hat{\varphi}_{\mbox{\scriptsize SQ}}/2.
\end{eqnarray}
We then obtain
\begin{eqnarray}
\hat{U}_{\mbox{\scriptsize SQ}}
&\simeq&
-2E_{\mbox{\scriptsize J}}
\left[
\cos
\left(
\pi
\frac{\Phi_{\mbox{\scriptsize DC}}}{\Phi_{0}}
\right)
-
\frac{\pi}{\Phi_{0}}
\sin
\left(
\pi
\frac{\Phi_{\mbox{\scriptsize DC}}}{\Phi_{0}}
\right)
\hat{\Phi}_{q}
\right] \nonumber \\
&&
\times
\left[
\cos
\left(
\frac{\varphi_{\mbox{\scriptsize DC}}}{2}
\right)
-
\frac{1}{2}
\sin
\left(
\frac{\varphi_{\mbox{\scriptsize DC}}}{2}
\right)
\hat{\varphi}_{\mbox{\scriptsize SQ}}
\right.
\nonumber \\
&&
\left.
-
\frac{1}{8}
\cos
\left(
\frac{\varphi_{\mbox{\scriptsize DC}}}{2}
\right)
\hat{\varphi}_{\mbox{\scriptsize SQ}}^{2}
\right].
\end{eqnarray}

Regarding the dc SQUID ring as an LC circuit, let $L_{\mbox{\scriptsize J}}$ denote the Josephson inductance of the SQUID and $C_{\mbox{\scriptsize SQ}}$ the total SQUID capacitance.
By analogy with the quantization of a harmonic oscillator with angular frequency
$\omega=(L_{\mbox{\scriptsize J}}C_{\mbox{\scriptsize SQ}})^{-1/2}$,
the SQUID mode can be expressed in terms of the creation and annihilation operators as
\begin{equation}
\hat{\varphi}_{\mbox{\scriptsize SQ}}
=
\frac{4\pi}{\Phi_{0}}
\sqrt{\frac{\hbar\omega L_{\mbox{\scriptsize J}}}{2}}
(\hat{a}+\hat{a}^{\dagger}).
\end{equation}

Here, we consider how to choose the bias current of the dc SQUID.
Since $\langle\hat{\Phi}_{q}\rangle=0$, we have
$\langle\hat{\Phi}_{\mbox{\scriptsize tot}}\rangle=\Phi_{\mbox{\scriptsize DC}}$.
From Eq.~(\ref{quantization-condition-flux-qubit-1}), the magnetic flux threading the inner ring in Fig.~\ref{figure01} is $(n-f)\Phi_{0}$, and, with $n$ an integer, $f=1/2$ is a suitable choice for making the energy levels of $|L\rangle$ and $|R\rangle$ equal.
However, this condition does not apply to the magnetic flux $\Phi_{\mbox{\scriptsize DC}}$ threading the outer ring in Fig.~\ref{figure01}.
In fact, the contribution of the magnetic flux threading the region between the two rings allows us to have
$\Phi_{\mbox{\scriptsize DC}}/\Phi_{0}\neq n-f$.
Thus, even when $f=1/2$, it is possible to avoid
$\Phi_{\mbox{\scriptsize DC}}=[n-(1/2)]\Phi_{0}$.
As can be seen from Eq.~(\ref{formula-L-J-1}) below, this prevents both $L_{\mbox{\scriptsize J}}$ from diverging and the angular frequency of the harmonic oscillator obtained by quantizing the dc SQUID from becoming zero.

Furthermore, since
$\langle\hat{\varphi}_{\mbox{\scriptsize SQ}}\rangle=0$,
we have
$\langle\hat{\varphi}_{\mbox{\scriptsize tot}}\rangle=\varphi_{\mbox{\scriptsize DC}}$.
Thus, we obtain
\begin{eqnarray}
I_{\mbox{\scriptsize B}}
&=&
\langle
\hat{J}_{\mbox{\scriptsize tot}}
\rangle \nonumber \\
&=&
2I_{0}
\sin
\left(
\frac{\varphi_{\mbox{\scriptsize DC}}}{2}
\right)
\cos
\left(
\pi
\frac{\Phi_{\mbox{\scriptsize DC}}}{\Phi_{0}}
\right)
\end{eqnarray}
Here, setting $\theta=\varphi_{\mbox{\scriptsize DC}}/2$, we regard $I_{\mbox{\scriptsize B}}$ as a function of $\theta$ and write its variation as
$I_{\mbox{\scriptsize B}}(\theta+\delta\theta)=I_{\mbox{\scriptsize B}}(\theta)+\delta J$.
Then, we obtain
\begin{eqnarray}
L_{\mbox{\scriptsize J}}
&=&
\frac{[\Phi_{0}/(2\pi)]\delta\theta}{\delta J} \nonumber \\
&=&
\frac{\Phi_{0}}
{4\pi I_{0}
\cos
(
\pi
\Phi_{\mbox{\scriptsize DC}}/\Phi_{0}
)
\cos
(
\varphi_{\mbox{\scriptsize DC}}/2
)
},
\label{formula-L-J-1}
\end{eqnarray}
The condition for the existence of a stable harmonic oscillator mode is
\begin{equation}
\cos
\left(
\pi
\frac{\Phi_{\mbox{\scriptsize DC}}}{\Phi_{0}}
\right)
\cos
\left(
\frac{\varphi_{\mbox{\scriptsize DC}}}{2}
\right)
>0,
\end{equation}
because $\omega=(L_{\mbox{\scriptsize J}}C_{\mbox{\scriptsize SQ}})^{-1/2}$
and $L_{\mbox{\scriptsize J}}$ must have a positive, finite value.
Therefore, we must have $-\pi<\varphi_{\mbox{\scriptsize DC}}<\pi$.
In particular, note that $\varphi_{\mbox{\scriptsize DC}}\neq\pm\pi$.

From Eq.~(\ref{U-SQ-1}), we see that when
\begin{equation}
\left|
\frac{1}{2}
\sin
\left(
\frac{\varphi_{\mbox{\scriptsize DC}}}{2}
\right)
\right|
>
\left|
\frac{1}{8}
\cos
\left(
\frac{\varphi_{\mbox{\scriptsize DC}}}{2}
\right)
\right|,
\end{equation}
the term proportional to
$\hat{\varphi}_{\mbox{\scriptsize SQ}}$
dominates over the term proportional to
$\hat{\varphi}_{\mbox{\scriptsize SQ}}^{2}$.
A somewhat lengthy calculation shows that, for example, when $0.490<\varphi_{\mbox{\scriptsize DC}}<\pi$,
the term proportional to
$\hat{\varphi}_{\mbox{\scriptsize SQ}}$
is dominant.
On the other hand, for example, when $\varphi_{\mbox{\scriptsize DC}}=0$,
the term proportional to
$\hat{\varphi}_{\mbox{\scriptsize SQ}}^{2}$
is dominant.

Here, we consider the case where the term proportional to $\hat{\varphi}_{\mbox{\scriptsize SQ}}$ is dominant.
Retaining only the dominant terms in $\hat{U}_{\mbox{\scriptsize SQ}}$, we obtain
\begin{eqnarray}
\hat{U}_{\mbox{\scriptsize JC}}
&=&
-4E_{\mbox{\scriptsize J}}
\left(
\frac{\pi}{\Phi_{0}}
\right)^{2}
\sin
\left(\pi
\frac{\Phi_{\mbox{\scriptsize DC}}}{\Phi_{0}}
\right)
\sin
\left(
\frac{\varphi_{\mbox{\scriptsize DC}}}{2}
\right) \nonumber \\
&&
\times
MI_{\mbox{\scriptsize P}}
\sqrt{\frac{\hbar\omega L_{\mbox{\scriptsize J}}}{2}}
(\hat{a}+\hat{a}^{\dagger})\hat{\sigma}_{z}.
\label{Hamiltonian-single-photon-JCM-1}
\end{eqnarray}
Recall that $|L\rangle$ and $|R\rangle$ represent
persistent currents flowing in opposite directions.
At the symmetry point $f=1/2$, tunneling between
the two wells gives the approximate energy eigenstates
$|0\rangle$ and $|1\rangle$ defined by Eq.~(\ref{LR-01-transformation-1}).
This is a change of basis, rather than a physical
operation applied to the qubit.
In this energy eigenbasis, the operator previously
denoted by $\hat{\sigma}_{z}$ in
$\hat{U}_{\mbox{\scriptsize JC}}$ is represented
by $\hat{\sigma}_{x}$.
Near resonance and for sufficiently weak coupling,
the rotating-wave approximation (RWA) yields
the interaction Hamiltonian of the single-photon JCM.

Next, we consider the case $\varphi_{\mbox{\scriptsize DC}}=0$.
Retaining only the dominant terms in $\hat{U}_{\mbox{\scriptsize SQ}}$, we obtain
\begin{eqnarray}
\hat{U}_{\mbox{\scriptsize TPR}}
&=&
-4E_{\mbox{\scriptsize J}}
\left(
\frac{\pi}{\Phi_{0}}
\right)^{3}
\sin
\left(\pi
\frac{\Phi_{\mbox{\scriptsize DC}}}{\Phi_{0}}
\right)
MI_{\mbox{\scriptsize P}}
\frac{\hbar\omega L_{\mbox{\scriptsize J}}}{2} \nonumber \\
&&
\times
(\hat{a}+\hat{a}^{\dagger})^{2}\hat{\sigma}_{z}.
\label{Hamiltonian-two-photon-JCM-1}
\end{eqnarray}
Therefore, for example, when $I_{\mbox{\scriptsize B}}=0$, $\varphi_{\mbox{\scriptsize DC}}=0$,
and $0<\Phi_{\mbox{\scriptsize DC}}/\Phi_{0}<1/2$, we obtain the two-photon Rabi model.
Replacing $\hat{\sigma}_{z}$ with $\hat{\sigma}_{x}$ and expanding the qubit-photon interaction term in $\hat{U}_{\mbox{\scriptsize TPR}}$, we obtain
\begin{equation}
[\hat{\sigma}_{+}\hat{a}^{2}+\hat{\sigma}_{-}(\hat{a}^{\dagger})^{2}]
+
[\hat{\sigma}_{+}(\hat{a}^{\dagger})^{2}+\hat{\sigma}_{-}\hat{a}^{2}]
+
\hat{\sigma}_{x}(2\hat{a}^{\dagger}\hat{a}+1).
\end{equation}
The first, second, and third terms rotate with angular frequencies,
$\omega_{0}-2\omega$, $\omega_{0}+2\omega$, and $\omega_{0}$, respectively.
Thus, if we set the resonance condition,
$\Delta=-\omega_{0}+2\omega=0$,
we can neglect the second and third terms because of the RWA and attain the Hamiltonian of the two-photon JCM.

Here, we check the values of the physical parameters that can be obtained experimentally.
If we write the interaction term of Eq.~(\ref{Hamiltonian-two-photon-JCM-1}) as
\begin{equation}
\hat{U}_{\mbox{\scriptsize TPR}}
=
-\hbar g_{2}\hat{\sigma}_{z}(\hat{a}+\hat{a}^{\dagger})^{2},
\label{interaction-term-TPR-0}
\end{equation}
we obtain
\begin{equation}
g_{2}
=
\frac{4E_{\mbox{\scriptsize J}}}{\hbar}
\left(
\frac{\pi}{\Phi_{0}}
\right)^{3}
\sin
\left(
\pi
\frac{\Phi_{\mbox{\scriptsize DC}}}{\Phi_{0}}
\right)
MI_{\mbox{\scriptsize p}}
\frac{\hbar\omega L_{\mbox{\scriptsize J}}}{2}.
\end{equation}
Because we set $\varphi_{\mbox{\scriptsize DC}}=0$ in the derivation of Eq.~(\ref{Hamiltonian-two-photon-JCM-1}), we can rewrite Eq.~(\ref{formula-L-J-1}) as
\begin{equation}
L_{\mbox{\scriptsize J}}
=
\frac{\Phi_{0}}
{4\pi I_{0}
\cos
(
\pi
\Phi_{\mbox{\scriptsize DC}}/\Phi_{0}
)
},
\end{equation}
The critical current of the two Josephson junctions in the dc SQUID is given by $I_{0}$, so that we obtain
\begin{equation}
E_{\mbox{\scriptsize J}}
=
\frac{\Phi_{0}I_{0}}{2\pi}
=
\frac{\Phi_{0}^{2}}
{8\pi^{2}L_{\mbox{\scriptsize J}}\cos(\pi\Phi_{\mbox{\scriptsize DC}}/\Phi_{0})}.
\label{EJ-1}
\end{equation}
The interaction term $U_{\mbox{\scriptsize TPR}}$ defined by Eq.~(\ref{Hamiltonian-two-photon-JCM-1}) is dominant when $I_{\mbox{\scriptsize B}}=0$, $\varphi_{\mbox{\scriptsize DC}}=0$, and $0<\Phi_{\mbox{\scriptsize DC}}/\Phi_{0}<1/2$.

The authors of Ref.~\cite{Chiorescu2004} experimentally realized the single-photon Rabi model with a Josephson-junction circuit.
In their setup, the frequencies were set to
$\omega/(2\pi)=2.91$ [GHz] and
$\omega_{0}/(2\pi)=5.9$--$6.4$ [GHz],
corresponding to an off-resonant case.
Coincidentally, $\omega_{0}\simeq 2\omega$ holds in the
experiment of Ref.~\cite{Chiorescu2004}.
In our proposal, we assume that the qubit is operated
at its symmetry point and that its transition frequency
satisfies the two-photon resonance condition
$\omega_0=2\omega$.
At this point, $\hbar\omega_0$ is the tunnel splitting
between the two lowest qubit energy eigenstates,
rather than the junction plasma energy.
Therefore, the physical parameter values in
Ref.~\cite{Chiorescu2004} provide a useful reference
for rough estimates of the parameters in our model.

Reference~\cite{Chiorescu2004} also reported that $L_{\mbox{\scriptsize J}}\simeq 80$ [pH] for $I_{\mbox{\scriptsize B}}=0$,
$M\simeq 20$ [pH], $I_{\mbox{\scriptsize p}}\simeq 0.18$ [$\mu$A].
Assuming $\Phi_{\mbox{\scriptsize DC}}/\Phi_{0}=1/4$ and putting the above physical parameters together, we obtain
$E_{\mbox{\scriptsize J}}=9.57\times 10^{-22}$ [J] and $g_{2}/(2\pi)=3.98$ [MHz].
The experimental parameters used above are summarized in Table~\ref{Table1}.

\begin{table}
\begin{center}
\begin{tabular}
{
p{0.52\linewidth}
p{0.24\linewidth}
p{0.18\linewidth}
}
Physical quantity
& Experimental value
& References \\
\hline
$\omega/(2\pi)$: the
frequency
of the photons
in the
single-photon
Rabi model
& 2.91 [GHz]
& \cite{Chiorescu2004} \\
\hline
$L_{\mbox{\scriptsize J}}$: the Josephson inductance
of the dc SQUID for $I_{\mbox{\scriptsize B}}=0$
& 80 [pH]
& \cite{Chiorescu2004} \\
\hline
$M$: the qubit-SQUID mutual inductance
& 20 [pH]
& \cite{Chiorescu2004} \\
\hline
$I_{\mbox{\scriptsize p}}$: the persistent current
of the flux qubit
& $0.18$ [$\mu$A]
& \cite{Krantz2019} \\
\hline
$g_{1}/(2\pi)$: the coupling constant
of the single-photon Rabi model
& 0.2 [GHz]
& \cite{Chiorescu2004} \\
\hline
$A_{\Phi}^{2}$: the frequency-independent part
of the power spectrum of the $1/f$
flux noise
& $(1\mu\Phi_{0})^{2}/\mbox{Hz}$
& \cite{Yoshihara2006,Koch2007b,Bylander2011,Anton2012,Krantz2019} \\
\hline
$T_{1}$: the energy relaxation time
\\
of the flux qubit
& $6$--$20$ [$\mu$s]
& \cite{Yan2016} \\
\hline
$\kappa/(2\pi)$: the photon-loss rate
of the dc SQUID
& $10.2$ [kHz]
& \cite{Goppl2008} \\
\end{tabular}
\end{center}
\caption{List of experimental values
}
\label{Table1}
\end{table}

To derive the single-photon and two-photon JCMs, we apply the RWA to Eqs.~(\ref{Hamiltonian-single-photon-JCM-1}) and (\ref{Hamiltonian-two-photon-JCM-1}).
Here, the following point should be noted.
For the RWA to be valid, the coupling constants must be sufficiently small compared with the relevant bare frequencies $\omega_{0}$ and $\omega$.
The effective interaction strength of the interaction term
$-\hbar g_{1}\hat{\sigma}_{z}(\hat{a}+\hat{a}^{\dagger})$
in the single-photon Rabi model with the coherent state $|\alpha\rangle$ is
$[g_{1}/(2\pi)]\sqrt{n+1}\sim[g_{1}/(2\pi)]|\alpha|$,
whereas that of the interaction term
$-\hbar g_{2}\hat{\sigma}_{z}(\hat{a}+\hat{a}^{\dagger})^{2}$
in the two-photon Rabi model is
$[g_{2}/(2\pi)]\sqrt{(n+1)(n+2)}\sim[g_{2}/(2\pi)]|\alpha|^{2}$.
Substituting $\alpha=12.0$ into the values of $g_{1}$ and $g_{2}$ given in Table~\ref{Table1} and above, we obtain
$[g_{1}/(2\pi)]|\alpha|=2.4$ [GHz]
and
$[g_{2}/(2\pi)]|\alpha|^{2}=0.573$ [GHz].

Here, we review the RWA for the two-photon Rabi model.
The Hamiltonian $\hat{U}_{\mbox{\scriptsize TPR}}$ given by Eq.~(\ref{interaction-term-TPR-0})
can be separated into the resonant term
$-\hbar g_{2}[\hat{\sigma}_{+}\hat{a}^{2}+\hat{\sigma}_{-}(\hat{a}^{\dagger})^{2}]$,
the counter-rotating term
$-\hbar g_{2}[\hat{\sigma}_{+}(\hat{a}^{\dagger})^{2}+\hat{\sigma}_{-}\hat{a}^{2}]$,
and the remainder
$-\hbar g_{2}\hat{\sigma}_{x}(2\hat{a}^{\dagger}\hat{a}+1)$
with replacing $\hat{\sigma}_{z}$ with $\hat{\sigma}_{x}$.
Assuming $\omega/(2\pi)=2.91$ [GHz] and
$\omega_{0}=2\omega$,
the two-photon resonance condition is satisfied, and the resonant term varies slowly.
Meanwhile, the counter-rotating term and the remainder oscillate at high frequencies of
$(\omega_{0}+2\omega)/(2\pi)=4\omega/(2\pi)=11.64$ [GHz]
and $\omega_{0}/(2\pi)=2\omega/(2\pi)=5.82$ [GHz],
respectively.
Thus, the conditions for the RWA to hold are
$[g_{2}/(2\pi)]|\alpha|^{2}\ll(\omega_{0}+2\omega)/(2\pi)$
and
$[g_{2}/(2\pi)]|\alpha|^{2}\ll \omega_{0}/(2\pi)$.
In the present case, because of $[g_{2}/(2\pi)]|\alpha|^{2}=0.573$ [GHz] with $|\alpha|=12.0$,
these conditions are satisfied, and we can obtain the two-photon JCM.
However, if $|\alpha|$ is made too large, the RWA ceases to be valid.
Therefore, $|\alpha|$ cannot be increased without limit.

The dc SQUID oscillator serves as the bosonic mode
in the present model.
A switching readout that drives this same SQUID
into a finite-voltage state cannot be assumed
to preserve the generated oscillator state.
We therefore do not adopt such a switching readout
as a state-preserving measurement protocol.

A physical implementation would require a method
for stopping the interaction and measuring the qubit
while preserving the oscillator state,
or for transferring the state to a separate storage mode
before readout.
The design and analysis of these operations
are beyond the scope of the present work.

\section{\label{section-noise}Possible Sources of Noise in the Generation of the Cat States}
We evaluate the effects of three types of noise that may arise in the superconducting quantum circuit shown in Fig.~\ref{figure01}.

\subsection{\label{subsection-f-inverse-flux-noise}Random Magnetic Flux Noise with $1/f$ Fluctuations}
One of the sources of dissipation in a flux qubit is $1/f$ flux noise \cite{Koch2007a}.
The following empirical observations regarding $1/f$ flux noise have been obtained experimentally.
First, this noise causes pure dephasing in the flux qubit.
Its power spectrum is given by
\cite{Yoshihara2006,Koch2007b,Bylander2011,Anton2012,Krantz2019}
\begin{equation}
S_{\Phi}(\omega)
=
A_{\Phi}^{2}
\left(
\frac{2\pi\times 1 \mbox{Hz}}{\omega}
\right)^{\gamma_{\Phi}},
\label{power-spectrum-experimental-estimation-1}
\end{equation}
where $0.8\leq\gamma_{\Phi}\leq 1.0$ and $A_{\Phi}^{2}\approx(1 \mu \Phi_{0})^{2}/\mbox{Hz}$.
This relation holds over the range of $\omega/(2\pi)$ from mHz to GHz.

The microscopic origin of this $1/f$ magnetic flux noise is believed to be due to the stochastic flipping of unpaired spins (magnetic dipoles) that reside on the surfaces of the superconducting metals comprising the qubit \cite{Krantz2019}.
To test this conjecture experimentally, Ref.~\cite{Braumuller2020} investigated the dependence of $A_{\Phi}$ on the characteristic length scale of the dc SQUID.
Reference~\cite{Sendelbach2008} also performed experiments in which the temperature of the dc SQUID was varied to adjust the surface density of unpaired spins and observe the resulting changes in the noise, thereby obtaining evidence supporting this conjecture.
Reference~\cite{Rower2023} experimentally investigated how $1/f$ magnetic flux noise is affected by the magnetic field applied externally to the dc SQUID.
The results showed that the external magnetic field increased the spin-cluster size and suppressed the $1/f$ noise.
These findings support the model proposed in Ref.~\cite{Koch2007a}.

Because the noise causes pure dephasing, the corresponding perturbation term in the Hamiltonian takes the form
$
\hat{H}'(t)=-I_{\mbox{\scriptsize p}}\delta\Phi(t)\hat{\sigma}_{z},
$
where $I_{\mbox{\scriptsize p}}$ is the persistent current and $\delta\Phi(t)$ is a temporally fluctuating random magnetic flux.
However, the following point should be noted.
To obtain the two-photon JCM, we transformed the basis vectors of the Hamiltonian as in Eq.~(\ref{LR-01-transformation-1}) and applied the RWA.
Therefore, we should replace the operator $\hat{\sigma}_{z}$ with $\hat{\sigma}_{x}$ and write
$
\hat{H}'(t)=-I_{\mbox{\scriptsize p}}\delta\Phi(t)\hat{\sigma}_{x}
$
in the Schr{\"{o}}dinger picture.

\begin{figure}
\begin{center}
\includegraphics[width=1.0\linewidth]{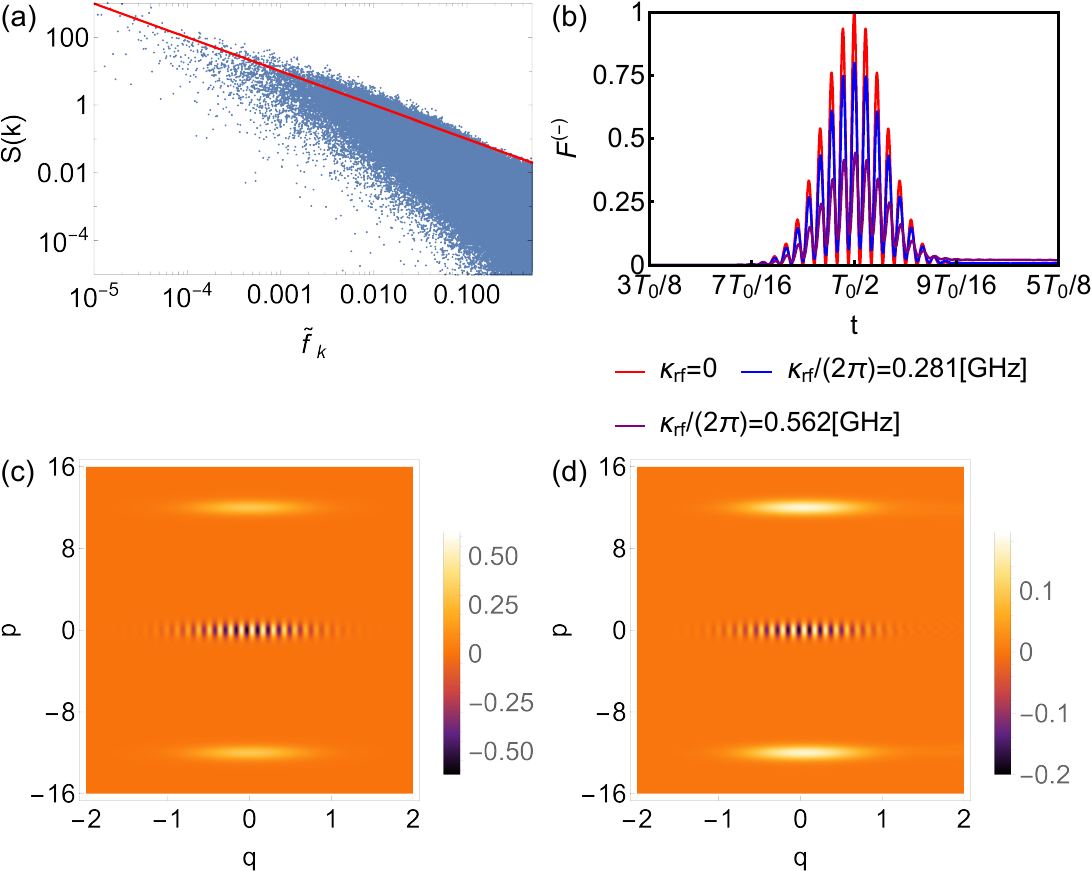}
\end{center}
\caption{
(a)
Spectral density of
$\{x_{k}: k=0, 1, 2, ..., 10^{6}-1\}$ generated by the recurrence relation for $x_{n}$ in Eq.~(\ref{one-dimensional_modified_Bernoulli_map-0}), plotted as a function of the dimensionless frequency $\tilde{f}_{k}$.
Specifically, $S(k)=|X_{k}|^{2}$.
The initial value is set to $x_{0}=-0.01$.
Both axes are shown on logarithmic scales.
The red straight line represents $S(k)=0.01\times \tilde{f}_{k}^{-1}$.
It agrees well with the upper envelope of the plotted distribution of $|X_{k}|^{2}$.
Thus, $\{x_{k}\}$ can be regarded as exhibiting $f^{-1}$ fluctuations.
(b)
$F^{(-)}$ plotted as a function of time $t$.
The parameters are set to
$g_{2}/(2\pi)=3.98$ [MHz],
$\omega/(2\pi)=2.91$ [GHz],
$\omega_{0}=2\omega$, and $\alpha=12.0$.
The red, blue, and purple curves correspond to $\kappa_{\mbox{\scriptsize rf}}/(2\pi)=0$, $0.281$, and $0.562$ [GHz], respectively.
(c), (d)
Heat maps of the Wigner function $W(q,p)$ of the approximate cat state obtained at $t=T_{0}(2)/2$.
Panels (c) and (d) correspond to
$\kappa_{\mbox{\scriptsize rf}}/(2\pi)=0$ and $0.562$ [GHz], respectively.
Although the plots in panels (c) and (d) look very similar, the color scale in panel (c) ranges from $-0.62$ to $+0.62$, whereas that in panel (d) ranges from $-0.201$ to $+0.195$.
Thus, the magnitudes of the positive and negative amplitudes differ substantially between the two panels.
Even in panel (d), where dissipation is present, the Wigner function takes negative values, indicating that non-Gaussianity persists.
Along the horizontal line $p=0$, the value of the Wigner function oscillates between positive and negative values.
The fidelity $F^{(-)}$ and the Wigner function $W(q,p)$ shown here are obtained in the interaction picture defined in Eq.~(\ref{interaction-picture-Hamiltonians-0}).
The same applies to the physical quantities shown in Figs.~\ref{figure04}, \ref{figure05}, \ref{figure06}, \ref{figure07}, \ref{figure08}, \ref{figure09}, \ref{figure10}, \ref{figure11}.
}
\label{figure03}
\end{figure}

We investigate by numerical simulation how the $1/f$ flux noise affects cat-state generation using the two-photon JCM.
To this end, we must first generate a random-number sequence that produces noise with $1/f$ fluctuations.
One method for generating such a random-number sequence $\{x_{0},x_{1},x_{2},...\}$ is the one-dimensional modified Bernoulli map with $B=2$ \cite{Aizawa1984a,Aizawa1984b,Shinkai2006}.
This method generates the random-number sequence according to the following recurrence relations:
\begin{eqnarray}
x_{n}
&=&
z_{n}-\frac{1}{2}, \nonumber \\
z_{n+1}
&=&
\left\{
\begin{array}{ll}
z_{n}+2z_{n}^{2} & (0\leq z_{n}\leq 0.5), \\
z_{n}-2(1-z_{n})^{2} & (0.5<z_{n}\leq 1). \\
\end{array}
\right.
\label{one-dimensional_modified_Bernoulli_map-0}
\end{eqnarray}
Let $\{X_{k}:k=0,1,...,N-1\}$ be the discrete Fourier transform of $\{x_{n}:n=0,1,...,N-1\}$, defined by
\begin{equation}
X_{k}
=
\frac{1}{\sqrt{N}}
\sum_{n=0}^{N-1}x_{n}e^{2\pi ink/N}.
\end{equation}
We also define the dimensionless frequency as $\tilde{f}_{k}=k/N$.
As shown in Fig.~\ref{figure03}(a),
Numerical calculations confirm that the following relation holds:
\begin{equation}
|X_{k}|^{2}\simeq 0.01 \times\tilde{f}_{k}^{-1}.
\label{discrete-Fourier-transfom-approximation-0}
\end{equation}

For the numerical simulation, we discretize the time-dependent interaction term $\hat{H}'(t)$ with respect to time as follows:
\begin{eqnarray}
t
&=&
t_{0},t_{1},t_{2},...,t_{N-1}, \nonumber \\
t_{n}
&=&
n\delta t,
\quad
t_{N-1}=T,
\quad
T=T_{0}(2)/2=\pi/(2g_{2}), \nonumber \\
\hat{H}'(t_{n})
&=&
-I_{\mbox{\scriptsize p}}\delta\Phi_{n}\hat{\sigma}_{z},
\quad
\delta\Phi_{n}=\delta\Phi(t_{n}).
\end{eqnarray}
Here, we have introduced the quantity $T_{0}(2)=\pi/g_{2}$.
Its physical meaning is given by Eq.~(\ref{l-photon-JCM_period-1}).
We now consider the discrete Fourier transform of
$\{\delta\Phi_{0},\delta\Phi_{1},...,\delta\Phi_{N-1}\}$:
\begin{equation}
\tilde{\Phi}_{k}
=
\sum_{n=0}^{N-1}\delta\Phi_{n}e^{-2\pi i kn/N}.
\end{equation}
We define the spectral density $S_{\Phi}(\omega)$ as
\begin{equation}
S_{\Phi}(\omega)
=
\int_{-\infty}^{\infty}dt\,e^{i\omega t}\langle\delta\Phi(t)\delta\Phi(0)\rangle.
\end{equation}
Discretizing this expression gives
\begin{equation}
S_{\Phi}(f_{k})
\simeq
\delta t\sum_{n=0}^{N-1}e^{i\omega_{k}t_{n}}
\langle\delta\Phi_{n}\delta\Phi_{0}\rangle,
\end{equation}
where $\omega_{k}=2\pi k/(N\delta t)$.
Therefore, we obtain
\begin{equation}
S_{\Phi}(f_{k})
=
\delta t
\sum_{n=0}^{N-1}e^{2\pi i kn/N}
\langle\delta\Phi_{n}\delta\Phi_{0}\rangle.
\end{equation}

Here, from an assumption that $\langle\delta\Phi_{n}\delta\Phi_{m}\rangle$ depends only on $(n-m)$, the Wiener-Khinchin theorem gives
\begin{equation}
S_{\Phi}(f_{k})
=
\frac{\delta t}{N}\langle|\tilde{\Phi}_{k}|^{2}\rangle.
\end{equation}
From Eq.~(\ref{power-spectrum-experimental-estimation-1}), we obtain
\begin{equation}
S_{\Phi}(f_{k})
=
\frac{A_{\Phi}^{2}}{f_{k}}
=
\frac{N\delta t}{k}A_{\Phi}^{2}
=
\frac{\delta t}{N}|\tilde{\Phi}_{k}|^{2},
\end{equation}
where $f_{k}=k/(N\delta t)$.

Setting $\delta\Phi_{n}=c_{\mbox{\scriptsize rf}}x_{n}$, we obtain
\begin{eqnarray}
\tilde{\Phi}_{k}
&=&
\sum_{n=0}^{N-1}\delta\Phi_{n}e^{2\pi i kn/N} \nonumber \\
&=&
c_{\mbox{\scriptsize rf}}\sum_{n=0}^{N-1}x_{n}e^{2\pi i kn/N} \nonumber \\
&=&
c_{\mbox{\scriptsize rf}}\sqrt{N}X_{k},
\end{eqnarray}
and
\begin{equation}
S_{\Phi}(f_{k})
=
\frac{\delta t}{N}|\tilde{\Phi}_{k}|^{2}
=
\delta tc_{\mbox{\scriptsize rf}}^{2}|X_{k}|^{2}.
\end{equation}
Therefore,
\begin{equation}
\frac{A_{\Phi}^{2}}{f_{k}}
=
\delta tc_{\mbox{\scriptsize rf}}^{2}|X_{k}|^{2}.
\end{equation}
From Eq.~(\ref{discrete-Fourier-transfom-approximation-0}), we obtain
\begin{equation}
\frac{A_{\Phi}^{2}}{f_{k}}
\simeq
0.01\times\frac{\delta tc_{\mbox{\scriptsize rf}}^{2}}{\tilde{f}_{k}}.
\end{equation}
Recalling that $\tilde{f}_{k}=k/N=f_{k}\delta t$, we obtain $A_{\Phi}^{2}=0.01\times c_{\mbox{\scriptsize rf}}^{2}$,
or equivalently, $c_{\mbox{\scriptsize rf}}=10.0\times A_{\Phi}$.

Adding $\hat{H}'(t)$ as a perturbation with the interaction picture to the principal Hamiltonian term $\hat{C}_{2}$, we obtain
\begin{eqnarray}
&&
\hat{C}_{2}+\hat{H}'(t) \nonumber \\
&=&
-\frac{\hbar}{2}\Delta\hat{\sigma}_{z}+\hbar g_{2}[\hat{\sigma}_{+}\hat{a}^{2}+\hat{\sigma}_{-}(\hat{a}^{\dagger})^{2}] \nonumber \\
&&
-I_{\mbox{\scriptsize p}}\delta\Phi(t)[\cos(2\omega t)\hat{\sigma}_{x}-\sin(2\omega t)\hat{\sigma}_{y}],
\label{Hamiltonian-with-f-inverse-noise-1}
\end{eqnarray}
where we use a formula,
\begin{eqnarray}
&&
\exp(i\hat{C}_{1}t/\hbar)\hat{\sigma}_{x}\exp(-i\hat{C}_{1}t/\hbar) \nonumber \\
&=&
\exp(i\omega t\hat{\sigma}_{z})\hat{\sigma}_{x}\exp(-i\omega t\hat{\sigma}_{z}) \nonumber \\
&=&
\cos(2\omega t)\hat{\sigma}_{x}
-
\sin(2\omega t)\hat{\sigma}_{y}.
\end{eqnarray}
We now consider the resonant case, i.e., $\Delta=-\omega_{0}+2\omega=0$.
It is difficult to analytically describe the time evolution of the state vector under the Hamiltonian in Eq.~(\ref{Hamiltonian-with-f-inverse-noise-1}) in closed form.
Therefore, we investigate the time evolution of the density matrix using the following numerical method.

First, we take the initial photon wave function to be a coherent state with $\alpha=12.0$.
Next, we truncate the photon number-state basis to the finite set
$\{|n\rangle_{\mbox{\scriptsize P}}: n=0, 1, ..., 300\}$.
The mean photon number of the coherent state $|\alpha\rangle$ is $|\alpha|^{2}=144$, and this finite set of photon number states is sufficient for reliable numerical calculations.
Since the atomic basis consists of the two vectors $|g\rangle_{\mbox{\scriptsize A}}$ and $|e\rangle_{\mbox{\scriptsize A}}$,
$\hat{C}_{2}+\hat{H}'(t)$ is represented by a $602\times602$ matrix in this approximation.
We then approximate the unitary time-evolution operator $\hat{U}(n\Delta t)$ at time $n\Delta t$ by
\begin{equation}
\hat{U}'(n\Delta t)=
\sum_{m=0}^{4}
\frac{1}{m!}
\left[
(-\frac{i}{\hbar}\Delta t)
(\hat{C}_{2}+\hat{H}'(n\Delta t))
\right]
^{m},
\end{equation}
where the time variable $t$ is discretized as $t_{n}=n\Delta t$, $n=0, 1, ..., N$.
We set $N=2\times 10^{5}$ and $\Delta t=T_{0}(2)/N$.

When evaluating $\hat{H}'(n\Delta t)$, we set $\delta\Phi(n\Delta t)=c_{\mbox{\scriptsize rf}}x_{n}$.
The state of the atom and photon at time $t_{n}$ is given by
\begin{equation}
|\Psi(n\Delta t)\rangle
=
\prod_{k=0}^{n}
\hat{U}'(k\Delta t)|g\rangle_{\mbox{\scriptsize A}}|\alpha\rangle_{\mbox{\scriptsize P}},
\end{equation}
where $|\alpha\rangle_{\mbox{\scriptsize P}}$ in the above equation is the coherent state truncated to $\{|n\rangle_{\mbox{\scriptsize P}}: n=0, 1, ..., 300\}$.
The density matrix is obtained as $\hat{\rho}(n\Delta t)=|\Psi(n\Delta t)\rangle\langle\Psi(n\Delta t)|$.

Here, we perform $10^{3}$ trials in a Monte Carlo simulation and average $\hat{\rho}(n\Delta t)$ over these trials.
We then calculate the following physical quantities using the resulting averaged density matrix $\rho(t)$.
Let $P_{g}(t)$ denote the probability of detecting $|g\rangle_{\mbox{\scriptsize A}}$ at time $t$:
\begin{equation}
P_{g}(t)
=
\mbox{Tr}_{\mbox{\scriptsize P}}[_{\mbox{\scriptsize A}}\langle g|\hat{\rho}(t)|g\rangle_{\mbox{\scriptsize A}}].
\label{definition-Prob-g-t-0}
\end{equation}
Furthermore, let $F^{(-)}(t)$ denote the fidelity with respect to the cat state of the photon state obtained through wave-function collapse upon detecting $|g\rangle_{\mbox{\scriptsize A}}$:
\begin{equation}
F^{(-)}(t)
=
\frac{1}{P_{g}(t)}
{}_{\mbox{\scriptsize A}}\langle g|
{}_{\mbox{\scriptsize P}}\langle\mbox{cat}^{(-)},\alpha|
\hat{\rho}(t)
|\mbox{cat}^{(-)},\alpha\rangle_{\mbox{\scriptsize P}}
|g\rangle_{\mbox{\scriptsize A}},
\label{definition-fidelity-F-minus-0}
\end{equation}
where
\begin{equation}
|\mbox{cat}^{(\pm)},\alpha\rangle_{\mbox{\scriptsize P}}
=
(1/\sqrt{2})
(|c\alpha\rangle_{\mbox{\scriptsize P}}\pm i|c^{*}\alpha\rangle_{\mbox{\scriptsize P}}),
\label{cat-states-definition-0}
\end{equation}
and $c=e^{-i\pi/2}$.
Defining $T_{0}(2)=\pi/g_{2}$, the cat state shown in Eq.~(\ref{2-mode-JCM-cat-state-obtained-1}) is generated at time $T_{0}(2)/2$.
The experimental values used in this section and the references from which they were obtained are summarized in Table~\ref{Table1}.

Defining $\kappa_{\mbox{\scriptsize rf}}=c_{\mbox{\scriptsize rf}}I_{\mbox{\scriptsize p}}/\hbar$
and setting $I_{\mbox{\scriptsize p}}=0.18$ [$\mu$A] according to Table~\ref{Table1}, we obtain
$\kappa_{\mbox{\scriptsize rf}}/(2\pi)=5.62$ [MHz].
Figure~\ref{figure03}(b) shows the time evolution of $F^{(-)}$ for
$\kappa_{\mbox{\scriptsize rf}}/(2\pi)=0$, $0.281$, and $0.562$ [GHz].
The figure shows that although $F^{(-)}$ reaches its maximum at $t=T_{0}(2)/2$, this maximum decreases as $\kappa_{\mbox{\scriptsize rf}}$ increases.
However, this graph indicates that almost no dissipation occurs for $\kappa_{\mbox{\scriptsize rf}}/(2\pi)=5.62$ [MHz].
At the qubit symmetry point, the flux-noise perturbation oscillates at $2\omega=\omega_0$ in the interaction picture [Eq.~(\ref{Hamiltonian-with-f-inverse-noise-1})]. These rapid oscillations suppress the first-order effect of low-frequency noise. Thus, even when $\kappa_{\mbox{\scriptsize rf}}$ is comparable to $g_2$, the fidelity reduction can remain small if both are much smaller than $2\omega$.

Figures~\ref{figure03}(c) and (d) show the Wigner functions $W(q,p)$ of the approximate cat states generated by detecting $|g\rangle_{\mbox{\scriptsize A}}$.
Figures~\ref{figure03}(c) and (d) correspond to $\kappa_{\mbox{\scriptsize rf}}/(2\pi)=0$ and $0.562$ [GHz], respectively.
The plots in Figs.~\ref{figure03}(c) and (d) use different color scales.
Thus, although the two plots look almost identical, the ranges of their positive and negative amplitudes differ substantially, indicating that Fig.~\ref{figure03}(d) represents a version of Fig.~\ref{figure03}(c) with reduced amplitudes.
Both plots contain regions with negative values, indicating that both states are non-Gaussian.
However, the smaller amplitudes in Fig.~\ref{figure03}(d) compared with Fig.~\ref{figure03}(c) clearly indicate the presence of dissipation.
Such dissipation occurs for $\kappa_{\mbox{\scriptsize rf}}/(2\pi)=0.562$ [GHz], whereas for the realistic value of $\kappa_{\mbox{\scriptsize rf}}/(2\pi)=5.62$ [MHz], the effects of this noise are almost negligible.
From these numerical results, we conclude that the random flux noise with the $1/f$ fluctuations does not pose a serious obstacle to cat-state generation.

\subsection{\label{subsection-T1-kappa-decay}$T_{1}$ Decay of the Flux Qubit and Photon-Loss Decay}

In this subsection, we consider the effects of energy relaxation noise that causes the $T_{1}$ decay of the flux qubit in the inner ring and the dissipation caused by photon loss in the bosonic mode obtained by quantizing the phase variable of the dc SQUID in the outer ring.
These two types of noise can be described by the following master equation:
\begin{eqnarray}
\frac{d}{dt}\hat{\rho}
&=&
-\frac{i}{\hbar}[\hat{C}_{2},\hat{\rho}]+\gamma_{1}D[\hat{\sigma}_{-}]\hat{\rho}
+\kappa D[\hat{a}]\hat{\rho}, \nonumber \\
D[\hat{L}]
&=&
\hat{L}\hat{\rho}\hat{L}^{\dagger}
-\frac{1}{2}(\hat{L}^{\dagger}\hat{L}\hat{\rho}+\hat{\rho}\hat{L}^{\dagger}\hat{L}),
\end{eqnarray}
where $\gamma_{1}=1/T_{1}$.

\begin{figure}
\begin{center}
\includegraphics[width=0.7\linewidth]{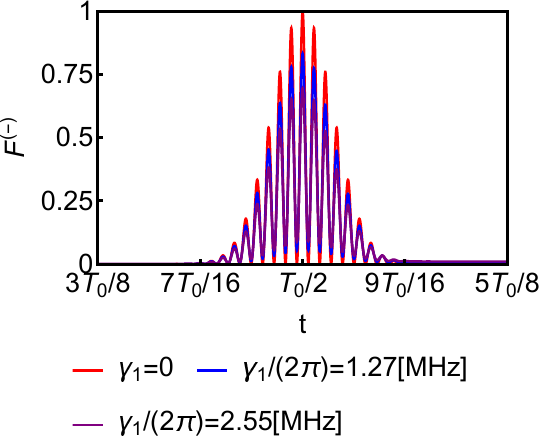}
\end{center}
\caption{
Time evolution of the fidelity $F^{(-)}$ of the photon state obtained by detecting $|g\rangle_{\mbox{\scriptsize A}}$ in the two-photon JCM with respect to the cat state $|\mbox{cat}^{(-)},\alpha\rangle_{\mbox{\scriptsize P}}$, taking into account the $T_{1}$ decay of the flux qubit.
The red, blue, and purple curves represent $T_{1}=\infty$, $125$, and $62.5$ [ns], respectively, corresponding to $\gamma_{1}/(2\pi)=0$, $1.27$, and $2.55$ [MHz].
Photon loss is assumed to be negligible, i.e., $\kappa=0$.
It can be seen that $F^{(-)}$ decreases as $\gamma_{1}$ increases.
The physical parameters are set to $g_{2}/(2\pi)=3.98$ [MHz], $\omega/(2\pi)=2.91$ [GHz], $\omega_{0}=2\omega$, $\Delta=0$, and $\alpha=12.0$.
}
\label{figure04}
\end{figure}

Experimental values of the energy relaxation time of the flux qubit have been reported in the range
$6\mbox{ [$\mu$s]}\leq T_{1}\leq 20\mbox{ [$\mu$s]}$ \cite{Yan2016}.
However, we numerically solve the master equation for $T_{1}=\infty$, $125$, and $62.5$ [ns], assuming $\kappa=0$, to examine the effects of the noise.
These values correspond to $\gamma_{1}/(2\pi)=0$, $1.27$, and $2.55$ [MHz].
Figure~\ref{figure04} shows $F^{(-)}$ plotted as a function of the time variable $t$ for $\alpha=12.0$, $g_{2}/(2\pi)=3.98$ [MHz],
$\Delta=0$, and $T_{1}=\infty$, $125$, and $62.5$ [ns].
As can be seen from Fig.~\ref{figure04}, the effect of relaxation on $F^{(-)}$ becomes appreciable only when $T_{1}$ is as short as approximately $125$ [ns].
In contrast, the experimentally demonstrated values of $T_{1}$ range from $6$ to $20$ [$\mu$s] and are therefore much longer.
We can thus conclude that the two-photon JCM is robust against $T_{1}$ decay.

\begin{figure}
\begin{center}
\includegraphics[width=1.0\linewidth]{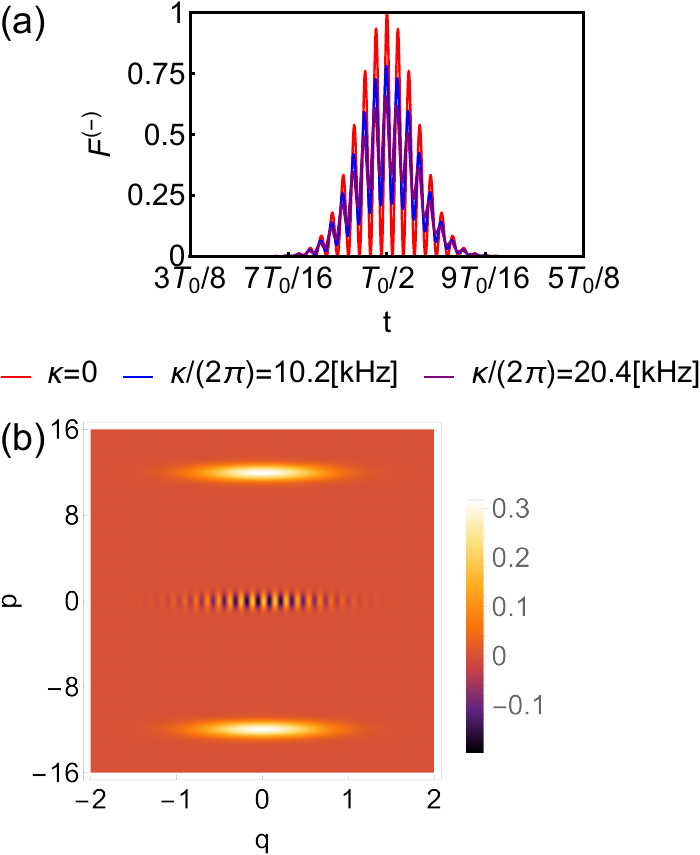}
\end{center}
\caption{
(a)
Fidelity $F^{(-)}$ of the photon state obtained by detecting $|g\rangle_{\mbox{\scriptsize A}}$ in the two-photon JCM with respect to the cat state $|\mbox{cat}^{(-)},\alpha\rangle$, plotted as a function of the time variable $t$ while taking into account the effect of photon loss.
The physical parameters are set to $g_{2}/(2\pi)=3.98$ [MHz], $\omega/(2\pi)=2.91$ [GHz], $\omega_{0}=2\omega$, $\Delta=0$, and $\alpha=12.0$.
The red, blue, and purple curves represent $\kappa/(2\pi)=0$, $10.2$, and $20.4$ [kHz], respectively.
It can be seen that the fidelity $F^{(-)}$ at $t=T_{0}(2)/2$ gradually decreases from unity as $\kappa$ increases.
(b)
The Wigner function $W(q,p)$ of the photon state obtained by detecting $|g\rangle_{\mbox{\scriptsize A}}$ at $t=T_{0}(2)/2$, plotted as a function of $q$ and $p$
for $\kappa/(2\pi)=20.4$ [kHz].
}
\label{figure05}
\end{figure}

To obtain a typical value of the photon-decay rate of the dc SQUID bosonic mode, we estimate it from the experimental values for the superconducting coplanar waveguide (CPW) resonator listed as ID L in Table II of Ref.~\cite{Goppl2008}: the measured resonance frequency $f_{0}=2.3464$ [GHz] and the measured quality factor $Q=2.3\times 10^{5}$.
These values give
$\kappa=2\pi f_{0}/Q=6.41\times 10^{4}$ [rad/s],
or equivalently, $\kappa/(2\pi)=10.2$ [kHz].
Here, $\kappa$ denotes the photon-number decay rate for the CPW resonator.
This value is used as a reference for our numerical study,
rather than as an experimentally established decay rate
of the dc SQUID mode.
Figure~\ref{figure05}(a) shows the time evolution of the fidelity $F^{(-)}$ for $\kappa/(2\pi)=0$, $10.2$, and $20.4$ [kHz], assuming $T_{1}=\infty$.
Figure~\ref{figure05}(b) shows the Wigner function $W(q,p)$ of the approximate cat state generated by detecting $|g\rangle$ for $\kappa/(2\pi)=20.4$ [kHz].
The experimental values used in this section and their source references are summarized in Table~\ref{Table1}.
Figure~\ref{figure05}(b) shows that non-Gaussianity persists despite dissipation at $\kappa/(2\pi)=20.4$ [kHz].
However, compared with the Wigner function shown in Fig.~\ref{figure03}(c), the plot in Fig.~\ref{figure05}(b) exhibits a smaller amplitude range between positive and negative values, clearly indicating the presence of dissipation.

For $\alpha=12.0$ and $\kappa/(2\pi)=20.4$ [kHz],
Fig.~\ref{figure05}(a) gives
$F^{(-)}\simeq0.617$ at $t=T_{0}(2)/2$,
assuming $T_{1}=\infty$.
Thus, we must conclude that the photon-loss decay is a serious obstacle for the generation of the cat states.
The calculations include photon loss during preparation
and assume that the interaction is switched off
instantaneously at the preparation time,
followed immediately by an ideal projective measurement
of the qubit.
Finite switching and readout times, measurement errors,
and additional disturbance of the oscillator
during these operations are not included.
The reported fidelities therefore characterize
the conditional states under these idealized assumptions,
rather than the output of a complete experimental sequence.

For the loss convention used here, the mode quality factor
is related to the photon-number decay rate by
$Q=\omega/\kappa$.
However, the conditional cat-state fidelity does not
in general follow the energy-decay law $\exp(-\kappa t)$.
We therefore do not infer a required quality factor
from the calculated fidelity.
Such an estimate would require a specified protocol
for stopping the interaction and reading out the qubit,
together with an evaluation of the state evolution
and measurement effects throughout the complete sequence.

\section{\label{section-time-evolution-phase}Time evolution of the phase of the cavity field in the multiphoton JCM}

As we can see from Eq.~(\ref{time-evolution-2}), when the initial state
$|g\rangle_{\mbox{\scriptsize A}}|\alpha\rangle_{\mbox{\scriptsize P}}$
is evolved under the Hamiltonian of the two-photon JCM, the cavity field splits into two components,
$|e^{-ig_{2}t}\alpha\rangle_{\mbox{\scriptsize P}}$
and
$|e^{ig_{2}t}\alpha\rangle_{\mbox{\scriptsize P}}$,
whose phases rotate in opposite directions.
To examine this behavior, we consider the Pegg-Barnett phase operator and evaluate the probability that its eigenstates appear in the cavity field. We then study how this probability evolves in time.

First of all, to define the Pegg-Barnett phase operator, we consider the following states of the cavity field:
\begin{equation}
|\phi_{m}\rangle_{s}
=
(s+1)^{-1/2}
\sum_{n=0}^{s}\exp(in\phi_{m})|n\rangle,
\end{equation}
where
\begin{equation}
\phi_{m}
=
\phi_{0}
+
\frac{2\pi}{s+1}m
\quad
\mbox{for $m=0, 1, 2, ..., s$,}
\end{equation}
and $\{|n\rangle: n=0, 1, ..., s\}$ are Fock number states.
The vectors $\{|\phi_{m}\rangle_{s}: m=0, 1, 2, ..., s\}$ form a complete orthonormal system in the $(s+1)$-dimensional Hilbert space.
Then, we can define the Pegg-Barnett phase operator as
\begin{equation}
\hat{\phi}
=
\sum_{m=0}^{s}\phi_{m}|\phi_{m}\rangle_{s}{}_{s}\langle\phi_{m}|.
\end{equation}

Next, we compute the probability that the cavity field has the phase $\phi_{m}$ in the $l$-photon JCM for $l=1,2,3$, and $4$
with setting the initial state of the atom and cavity photon as $|\Psi_{\mbox{\scriptsize I}}(0)\rangle=|g\rangle_{\mbox{\scriptsize A}}|\alpha\rangle_{\mbox{\scriptsize P}}$.
Then,
as shown in Eqs.~(\ref{interaction-picture-1}), (\ref{unitary-time-evolution-operator-1}), (\ref{unitary-time-evolution-operator-2}), and (\ref{unitary-time-evolution-operator-3}),
the state of the cavity field is given by
$\hat{\rho}_{\mbox{\scriptsize I, P}}(t)
=
\mbox{Tr}_{\mbox{\scriptsize A}}[|\Psi_{\mbox{\scriptsize I}}(t)\rangle\langle\Psi_{\mbox{\scriptsize I}}(t)|]$,
where
\begin{eqnarray}
|\Psi_{\mbox{\scriptsize I}}(t)\rangle
&=&
\hat{U}(t)|g\rangle_{\mbox{\scriptsize A}}|\alpha\rangle_{\mbox{\scriptsize P}} \nonumber \\
&=&
\left(
\begin{array}{cc}
\hat{u}_{00} & \hat{u}_{01}\\
\hat{u}_{10} & \hat{u}_{11}\\
\end{array}
\right)
\left(
\begin{array}{c}
0\\
|\alpha\rangle_{\mbox{\scriptsize P}}\\
\end{array}
\right).
\end{eqnarray}
The probability that the cavity field has a phase $\phi_{m}$ is described as
\begin{equation}
P(\phi_{m},t)
=
|_{s}\langle\phi_{m}|\hat{u}_{01}|\alpha\rangle_{\mbox{\scriptsize P}}|^{2}
+
|_{s}\langle\phi_{m}|\hat{u}_{11}|\alpha\rangle_{\mbox{\scriptsize P}}|^{2},
\end{equation}
where
\begin{eqnarray}
&&
{}_{s}\langle\phi_{m}|\hat{u}_{01}|\alpha\rangle_{\mbox{\scriptsize P}} \nonumber \\
&=&
-ig_{l}\alpha^{l}(s+1)^{-1/2}
\exp\left(-\frac{|\alpha|^{2}}{2}\right) \nonumber \\
&&
\times
\sum_{n=0}^{s}\exp(-in\phi_{m})\frac{\alpha^{n}}{\sqrt{n!}}
\frac{\sin(\sqrt{D_{n}^{(l)}}t)}{\sqrt{D_{n}^{(l)}}}, \nonumber \\
&&
{}_{s}\langle\phi_{m}|\hat{u}_{11}|\alpha\rangle_{\mbox{\scriptsize P}} \nonumber \\
&=&
(s+1)^{-1/2}
\exp\left(-\frac{|\alpha|^{2}}{2}\right) \nonumber \\
&&
\times
\sum_{n=0}^{s}\exp(-in\phi_{m})\frac{\alpha^{n}}{\sqrt{n!}} \nonumber \\
&&
\times
\left[
\cos(\sqrt{{D'}_{n}^{(l)}}t)
-i\frac{\Delta}{2}
\frac{\sin(\sqrt{{D'}_{n}^{(l)}}t)}{\sqrt{{D'}_{n}^{(l)}}}
\right],
\end{eqnarray}
\begin{eqnarray}
D_{n}^{(l)}
&=&
(\Delta/2)^{2}+g_{l}^{2}(n+l)!/n!, \nonumber \\
{D'}_{n}^{(l)}
&=&
(\Delta/2)^{2}+g_{l}^{2}[n!/(n-l)!]\Theta(n-l), \nonumber \\
\Theta(n)
&=&
\left\{
\begin{array}{ll}
1 & \mbox{for $n\geq 0$} \\
0 & \mbox{for $n<0$} \\
\end{array}
\right..
\end{eqnarray}

\begin{figure}
\begin{center}
\includegraphics[width=1.0\linewidth]{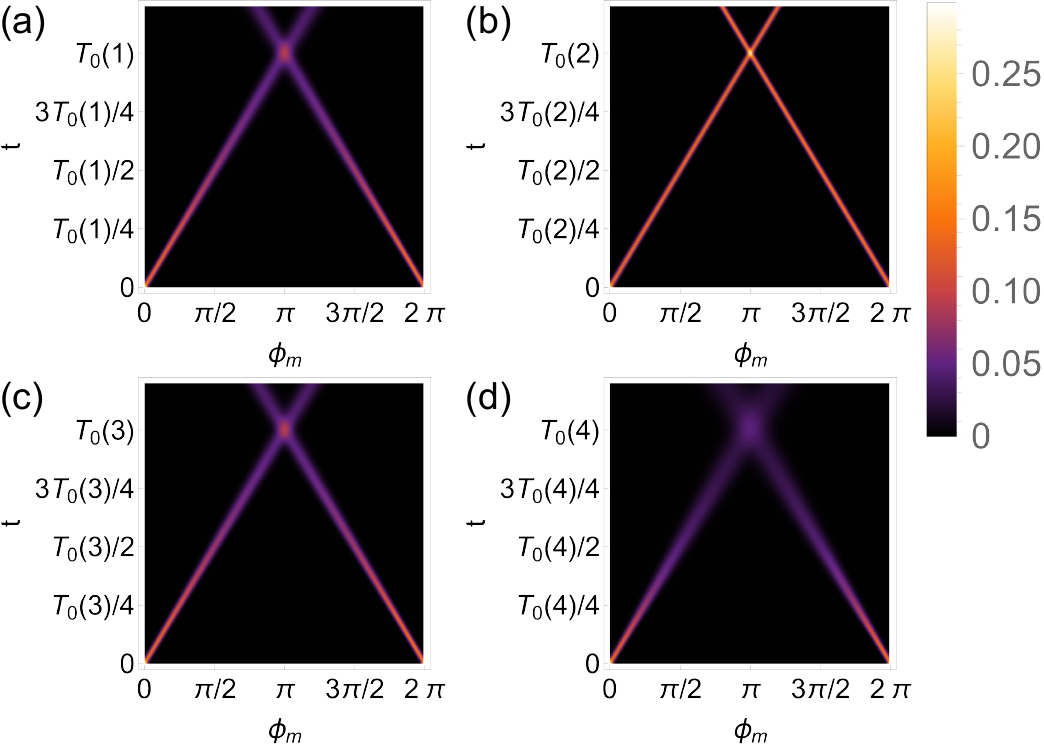}
\end{center}
\caption{2D plots of $P(\phi_{m},t)$ as functions of $\phi_{m}$ and $t$,
where
$\omega/(2\pi)=2.91$ [GHz],
$g_{l}/(2\pi)=3.98$ [MHz] for $l=1,2,3,4$,
$s=200$, $\alpha=12.0$, and $\phi_{0}=0$.
To set $\Delta=0$, we choose $\omega_{0}$ such that $\omega_{0}=l\omega$.
(a) $l=1$, $T_{0}(1)=3.02\times 10^{-6}$ [s],
(b) $l=2$, $T_{0}(2)=1.26\times 10^{-7}$ [s],
(c) $l=3$, $T_{0}(3)=6.98\times 10^{-9}$ [s],
(d) $l=4$, $T_{0}(4)=4.36\times 10^{-10}$ [s].}
\label{figure06}
\end{figure}

Figure~\ref{figure06} shows density plots of $P(\phi_{m},t)$ as functions of $\phi_{m}$ and $t$ for $l=1, 2, 3$ and $4$.
To impose $\Delta=0$, we choose $\omega_{0}=l\omega$.
Looking at Figs.~\ref{figure06}(a)-(d), we note that the local maxima of $P(\phi_{m},t)$ are located at $\phi_{m}=0$ at $t=0$, separate into two peaks, and merge at $\pi$ as time progresses.
Here, we pay attention to the period required for the phase to evolve from zero to $\pi$.
From numerical computations, for $|\alpha|\gg 1$ and $s\gg|\alpha|^{2}$, we find that this period is well approximated by
\begin{equation}
T_{0}(l)
=
\frac{2\pi}{g_{l}|\alpha|^{l-2}l}.
\label{l-photon-JCM_period-1}
\end{equation}
The period $T_{0}(l)$ was originally obtained as an approximate expression for the collapse-revival period of the Rabi oscillations of the atomic state \cite{Azuma2024}.

\begin{figure}
\begin{center}
\includegraphics[width=1.0\linewidth]{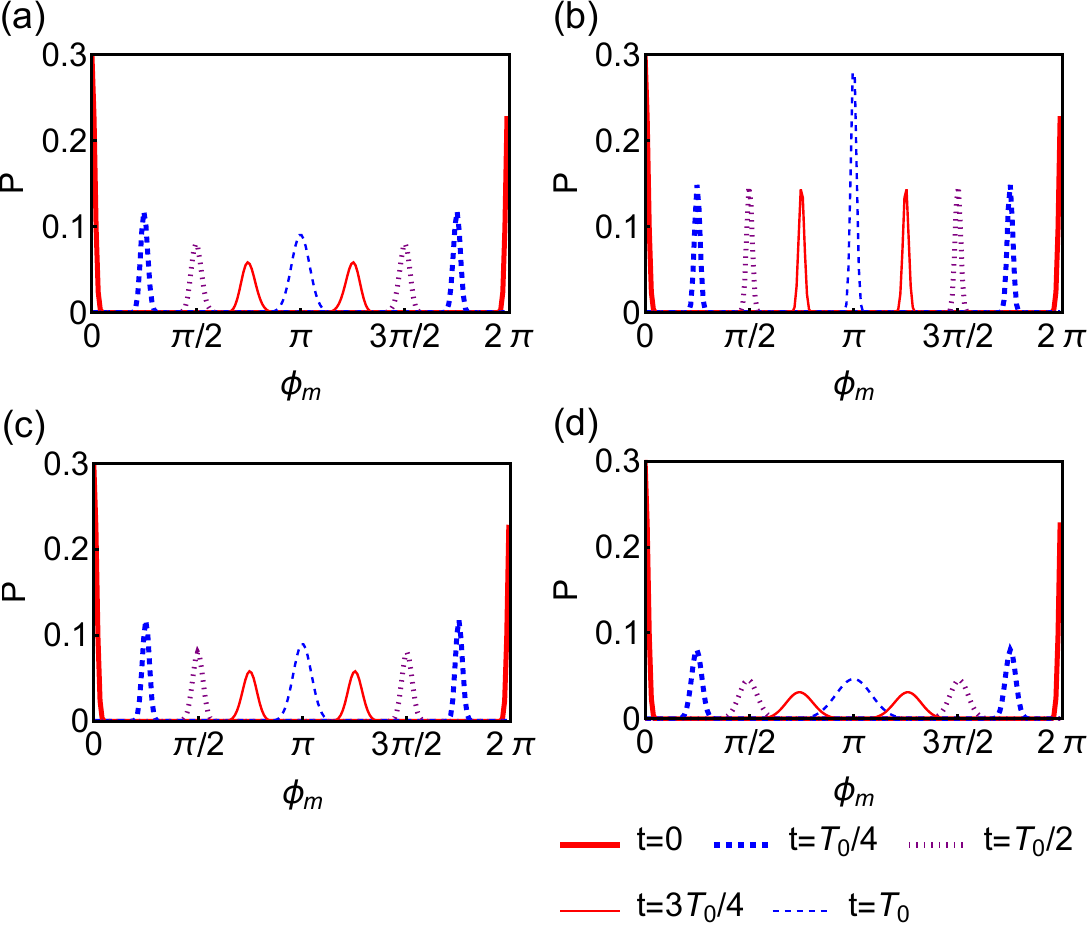}
\end{center}
\caption{Plots of $P(\phi_{m},t)$ as functions of $\phi_{m}$ for specific time $t$.
The thick solid red, thick dashed blue, thick dotted purple, thin solid red, and thin dashed blue curves represent $t=0$, $T_{0}(l)/4$, $T_{0}(l)/2$, $3T_{0}(l)/4$, and $T_{0}(l)$, respectively.
(a) $l=1$, (b) $l=2$, (c) $l=3$, (d) $l=4$.
The other parameters are set to the same values as those used in Fig.~\ref{figure06}(a)-(d).}
\label{figure07}
\end{figure}

Another observation from Figs.~\ref{figure06}(a)-(d) is as follows.
For $l=1, 3, 4$, the peaks of $P(\phi_{m},t)$ broaden as time progresses.
By contrast, for $l=2$, the local maxima of $P(\phi_{m},t)$ remain sharp even as time increases.
This observation is examined in more detail in Figs.~\ref{figure07}(a)-(d).
Figure~\ref{figure07} shows plots of probabilities $P(\phi_{m},t)$ as functions of the phase $\phi_{m}$ for specific time $t$.
Looking at Fig.~\ref{figure07}(a), (c), and (d), we note that the shape of the phase probability distribution becomes broader and less pronounced as time progresses for $l=1, 3$, and $4$.
By contrast, looking at Fig.~\ref{figure07}(b), we note that the shape of the phase distribution retains its sharpness even for increasing time for $l=2$.

As shown in Figs.~\ref{figure06} and \ref{figure07}, the Pegg-Barnett phase distribution illustrates the splitting and subsequent rotation of the two wave packets.
However, this does not guarantee that the cavity field is a coherent superposition of the two wave packets.
Even if these two wave packets form an incoherent mixture, it is still possible to obtain a phase distribution with two sharp peaks.
Nevertheless, the analysis in this section is useful for gaining a rough understanding of the time evolution of the cavity field.

\section{\label{section-temperature}Detection probability and fidelity of the approximately generated cat state when temperature is introduced into the system}

In this section, we examine how the detection probability and fidelity of the approximately generated cat state depend on temperature when thermal effects are introduced into the system.
To introduce temperature, we proceed as follows.
At the initial time $t=0$, the photons are brought into contact with a thermal bath at temperature $T$, while the atom is kept at zero temperature.
Under these conditions, the joint state of the atom and the photons is given by the following expression:
\begin{equation}
|g\rangle_{\mbox{\scriptsize A}}{}_{\mbox{\scriptsize A}}\langle g|
\otimes
\hat{\rho}_{\mbox{\scriptsize P,th}}(\beta),
\end{equation}
\begin{equation}
\hat{\rho}_{\mbox{\scriptsize P,th}}(\beta)
=
\sum_{n=0}^{\infty}p(n,\beta)
|n\rangle_{\mbox{\scriptsize P}}{}_{\mbox{\scriptsize P}}\langle n|,
\end{equation}
\begin{equation}
p(n,\beta)
=
[1-\exp(-\beta\hbar\omega)]\exp(-n\beta\hbar\omega),
\end{equation}
where $\beta=1/(k_{\mbox{\scriptsize B}}T)$,
$k_{\mbox{\scriptsize B}}$ is the Boltzmann constant,
$\omega$ is the photon angular frequency,
and
the function 
$p(n,\beta)$
is the probability distribution given by Bose-Einstein statistics.

Next, we apply the displacement operator
$\hat{D}(\alpha)=\exp(\alpha\hat{a}^{\dagger}-\alpha^{*}\hat{a})$
to the photons.
The state of the system then becomes
\begin{equation}
|g\rangle_{\mbox{\scriptsize A}}{}_{\mbox{\scriptsize A}}\langle g|
\otimes
\hat{D}(\alpha)\hat{\rho}_{\mbox{\scriptsize P,th}}(\beta)\hat{D}^{\dagger}(\alpha),
\label{initial-state-thermal-1}
\end{equation}
At this stage, the thermal bath is disconnected from the photons.
Thus, the initial state of the system is given by Eq.~(\ref{initial-state-thermal-1}).

Here, we introduce the following notation:
\begin{equation}
|\alpha,n\rangle_{\mbox{\scriptsize P}}
=
\hat{D}(\alpha)|n\rangle_{\mbox{\scriptsize P}},
\quad
|\alpha,0\rangle_{\mbox{\scriptsize P}}
=
|\alpha\rangle_{\mbox{\scriptsize P}}.
\end{equation}
Then, the initial state of the system can be written as follows:
\begin{equation}
\hat{\rho}(0)
=
\sum_{n=0}^{\infty}
p(n,\beta)
\left(
\begin{array}{cc}
0 & 0\\
0 & |\alpha,n\rangle_{\mbox{\scriptsize P}}{}_{\mbox{\scriptsize P}}\langle \alpha,n|\\
\end{array}
\right).
\end{equation}
Using the unitary operator
$\hat{U}(t)$ given in Eqs.~(\ref{unitary-time-evolution-operator-1}) and (\ref{unitary-time-evolution-operator-2}), the time evolution of the system can be written as
\begin{eqnarray}
&&
\hat{\rho}(t) \nonumber \\
&=&
\hat{U}(t)\hat{\rho}(0)\hat{U}^{\dagger}(t) \nonumber \\
&=&
\sum_{n=0}^{\infty}
p(n,\beta) \nonumber \\
&&
\times
\left(
\begin{array}{cc}
\hat{u}_{01} |\alpha,n\rangle_{\mbox{\scriptsize P}}{}_{\mbox{\scriptsize P}}\langle \alpha,n|\hat{u}_{01}^{\dagger}
&
\hat{u}_{01} |\alpha,n\rangle_{\mbox{\scriptsize P}}{}_{\mbox{\scriptsize P}}\langle \alpha,n|\hat{u}_{11}^{\dagger}
\\
\hat{u}_{11} |\alpha,n\rangle_{\mbox{\scriptsize P}}{}_{\mbox{\scriptsize P}}\langle \alpha,n|\hat{u}_{01}^{\dagger}
&
\hat{u}_{11} |\alpha,n\rangle_{\mbox{\scriptsize P}}{}_{\mbox{\scriptsize P}}\langle \alpha,n|\hat{u}_{11}^{\dagger}
\\
\end{array}
\right). \nonumber \\
&&
\end{eqnarray}

From Eq.~(\ref{definition-Prob-g-t-0}),
we obtain the following expressions for probabilities of detecting $|g\rangle_{\mbox{\scriptsize A}}$ and $|e\rangle_{\mbox{\scriptsize A}}$ at time $t$:
\begin{equation}
P_{g}(t)
=
\sum_{n=0}^{\infty}
p(n,\beta)
\langle \alpha,n|\hat{u}_{11}^{\dagger}\hat{u}_{11}|\alpha,n\rangle,
\end{equation}
\begin{equation}
P_{e}(t)
=
1-P_{g}(t).
\end{equation}
Thus, in order to compute $P_{g}(t)$ and $P_{e}(t)$ numerically, we need the quantity
$\langle \alpha,n|\hat{u}_{11}^{\dagger}\hat{u}_{11}|\alpha,n\rangle$.
A somewhat lengthy calculation yields the following result:
\begin{eqnarray}
&&
\langle \alpha,n|\hat{u}_{11}^{\dagger}\hat{u}_{11}|\alpha,n\rangle \nonumber \\
&=&
e^{-|\alpha|^{2}}
\sum_{j=0}^{\infty}
\sum_{j'=0}^{\infty}
\sum_{k=0}^{\infty}
\sum_{k'=0}^{\infty}
\frac{(\alpha^{*})^{j'}(-\alpha)^{k'}\alpha^{j}(-\alpha^{*})^{k}}{j!j'!k!k'!} \nonumber \\
&&
\times
f_{1}(n,k')f_{1}(n,k)f_{2}(n,k',j')f_{2}(n,k,j) \nonumber \\
&&
\times
\left[
\cos
\left(
\sqrt{{D'}_{n-k'+j'}^{(l)}}t
\right)
+
i
\frac{\Delta}{2}
\frac{
\sin
\left(
\sqrt{{D'}_{n-k'+j'}^{(l)}}t
\right)
}
{
\sqrt{{D'}_{n-k'+j'}^{(l)}}
}
\right] \nonumber \\
&&
\times
\left[
\cos
\left(
\sqrt{{D'}_{n-k+j}^{(l)}}t
\right)
-
i
\frac{\Delta}{2}
\frac{
\sin
\left(
\sqrt{{D'}_{n-k+j}^{(l)}}t
\right)
}
{
\sqrt{{D'}_{n-k+j}^{(l)}}
}
\right] \nonumber \\
&&
\times
\delta_{-k'+j',-k+j},
\end{eqnarray}
where
\begin{equation}
f_{1}(n,k)
=
\left\{
\begin{array}{ll}
\prod_{m=1}^{k}
\sqrt{n-k+m}
& \mbox{for $k\geq 1$,} \\
1 & \mbox{for $k=0$,} \\
\end{array}
\right.
\end{equation}
\begin{equation}
f_{2}(n,k,j)
=
\left\{
\begin{array}{ll}
\prod_{m=1}^{j}
\sqrt{n-k+m}
& \mbox{for $j\geq 1$,} \\
1 & \mbox{for $j=0$.} \\
\end{array}
\right.
\end{equation}

Next, for the state $\hat{\rho}(t)$, we evaluate the fidelity 
$F^{(-)}(t)$ of obtaining the cat state 
$|\mbox{cat}^{(-)},\alpha\rangle$ when $|g\rangle_{\mbox{\scriptsize A}}$ is detected, and the fidelity $F^{(+)}(t)$ of obtaining the cat state $|\mbox{cat}^{(+)},\alpha\rangle$ when 
$|e\rangle_{\mbox{\scriptsize A}}$ is detected.
These cat states are defined by Eq.~(\ref{cat-states-definition-0}).

Then, $F^{(-)}(t)$ given by Eq.~(\ref{definition-fidelity-F-minus-0}) can be written as follows:
\begin{equation}
F^{(-)}(t)
=
\frac{1}{2P_{g}(t)}
\sum_{n=0}^{\infty}
p(n,\beta)
|
(\langle c\alpha|+i\langle c^{*}\alpha|)
\hat{u}_{11}
|\alpha,n\rangle
|^{2}.
\end{equation}
A slightly tough calculation gives the following result:
\begin{eqnarray}
&&
(\langle c\alpha|+i\langle c^{*}\alpha|)
\hat{u}_{11}
|\alpha,n\rangle \nonumber \\
&=&
e^{-|\alpha|^{2}}
\sum_{j=0}^{\infty}
\sum_{k=0}^{n}
\frac{[(c^{*})^{n-k+j}+ic^{n-k+j}]|\alpha|^{2j}(\alpha^{*})^{n}(-1)^{k}}{\sqrt{(n-k+j)!}j!k!} \nonumber \\
&&
\times
f_{1}(n,k)f_{2}(n,k,j) \nonumber \\
&&
\times
\left[
\cos
\left(
\sqrt{{D'}_{n-k+j}^{(l)}}t
\right)
-
i
\frac{\Delta}{2}
\frac{
\sin
\left(
\sqrt{{D'}_{n-k+j}^{(l)}}t
\right)
}
{
\sqrt{{D'}_{n-k+j}^{(l)}}
}
\right]. \nonumber \\
&&
\end{eqnarray}
Thus, we can compute $F^{(-)}(t)$ numerically.

On the other hand, $F^{(+)}(t)$ can be written as follows:
\begin{eqnarray}
&&
F^{(+)}(t) \nonumber \\
&=&
\frac{1}{P_{e}(t)}
{}_{\mbox{\scriptsize A}}\langle e|
{}_{\mbox{\scriptsize P}}\langle\mbox{cat}^{(+)},\alpha|
\hat{\rho}(t)
|\mbox{cat}^{(+)},\alpha\rangle_{\mbox{\scriptsize P}}
|e\rangle_{\mbox{\scriptsize A}} \nonumber \\
&=&
\frac{1}{2P_{e}(t)}
\sum_{n=0}^{\infty}
p(n,\beta)
|
(\langle c\alpha|-i\langle c^{*}\alpha|)
\hat{u}_{01}
|\alpha,n\rangle
|^{2}.
\label{F-plus-t-1}
\end{eqnarray}
A straightforward but somewhat lengthy calculation gives the following result:
\begin{eqnarray}
&&
(\langle c\alpha|-i\langle c^{*}\alpha|)
\hat{u}_{01}
|\alpha,n\rangle \nonumber \\
&=&
e^{-|\alpha|^{2}} \nonumber \\
&&
\times
\sum_{j=0}^{\infty}
\sum_{k=0}^{n} \nonumber \\
&&
\times
\frac{[(c^{*})^{n-k+j-l}-ic^{n-k+j-l}]|\alpha|^{2j}(\alpha^{*})^{n-l}(-1)^{k}}{\sqrt{(n-k+j-l)!}j!k!} \nonumber \\
&&
\times
f_{1}(n,k)f_{2}(n,k,j)f_{3}(n,k,j,l) \nonumber \\
&&
\times
(-ig_{l})
\frac{
\sin
\left(
\sqrt{{D}_{n-k+j-l}^{(l)}}t
\right)
}
{
\sqrt{{D}_{n-k+j-l}^{(l)}}
},
\end{eqnarray}
where
\begin{equation}
f_{3}(n,k,j,l)
=
\Theta(n-k+j-l)
\prod_{m=1}^{l}
\sqrt{n-k+j-l+m},
\end{equation}
\begin{equation}
\Theta(n)
=
\left\{
\begin{array}{ll}
1 & \mbox{for $n\geq 0$,} \\
0 & \mbox{for $n<0$.} \\
\end{array}
\right.
\end{equation}
Thus, we can compute $F^{(+)}(t)$ numerically.

\begin{figure}
\begin{center}
\includegraphics[width=0.7\linewidth]{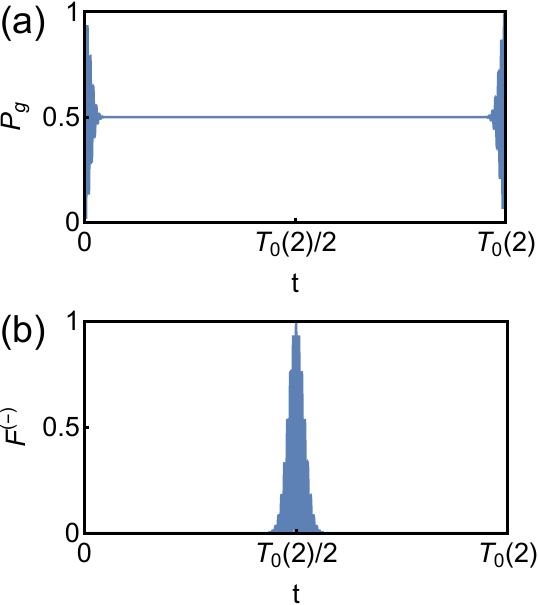}
\end{center}
\caption{
(a) Plot of $P_{g}(t)$ as a function of the time variable $t$.
(b) Plot of $F^{(-)}(t)$ as a function of the time variable $t$.
The results correspond to the two-photon JCM with $l=2$.
Since
$\omega/(2\pi)=2.91$ [GHz] and $\omega_{0}=2\omega$,
the system is in the resonant case with $\Delta=0$.
The parameters are $T=0.001$ [K],
$g_{2}/(2\pi)=3.98$ [MHz],
and $\alpha=12.0$.
The temporal period is
$T_{0}(2)=1.26\times 10^{-7}$ [s].
}
\label{figure08}
\end{figure}

The plots in Figs.~\ref{figure08}(a) and (b) show $P_{g}(t)$ and $F^{(-)}(t)$, respectively, as functions of the time variable $t$ for $l=2$.
The parameters $\omega$ and $\omega_{0}$ are chosen such that $\Delta=0$, corresponding to the resonant case.
From Fig.~\ref{figure08}(a), we see that the two-photon JCM exhibits collapse and revival of Rabi oscillations with the period $T_{0}(2)$.
Figure~\ref{figure08}(b) shows that $F^{(-)}$ reaches approximately unity at $t=T_{0}(2)/2$.
Thus, at $t=T_{0}(2)/2$, the cat state is obtained with a probability of about $1/2$.
These results in Figs.~\ref{figure08}(a) and (b) correspond to an extremely low temperature of $T=0.001$ [K], where effects of thermal noise are almost negligible.

\begin{figure}
\begin{center}
\includegraphics[width=1.0\linewidth]{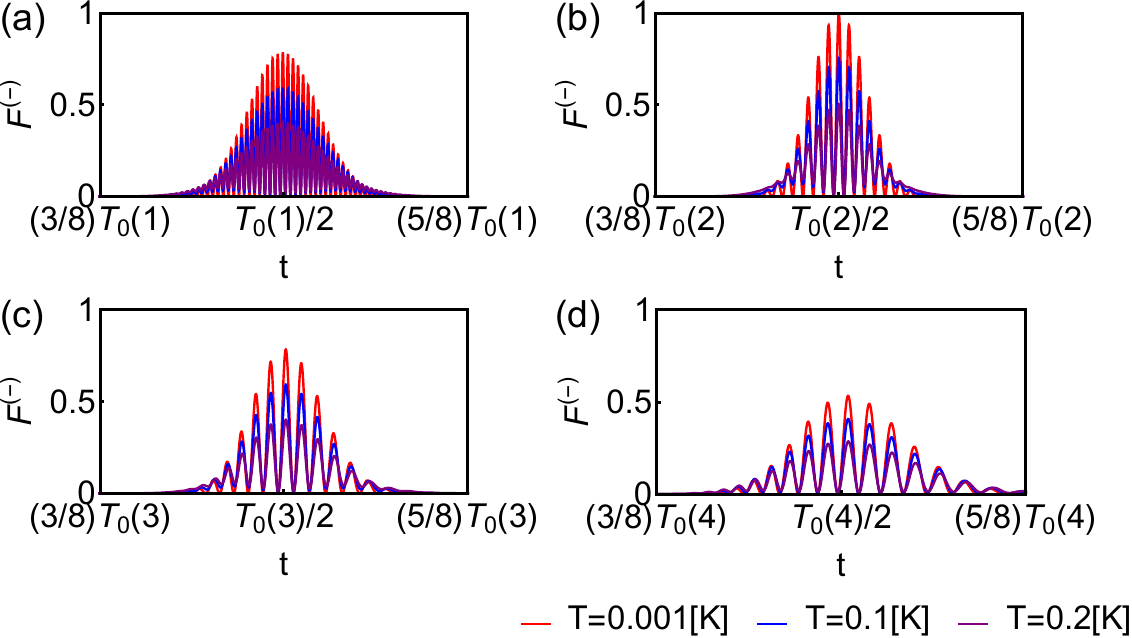}
\end{center}
\caption{
Plots of $F^{(-)}(t)$ as functions of the time variable $t$.
(a) $l=1$, $\omega_{0}=\omega$;
(b) $l=2$, $\omega_{0}=2\omega$;
(c) $l=3$, $\omega_{0}=3\omega$;
(d) $l=4$, $\omega_{0}=4\omega$.
With
$\omega/(2\pi)=2.91$ [GHz],
all cases (a)-(d) correspond to the resonant condition $\Delta=0$.
The parameters are
$g_{l}/(2\pi)=3.98$ [MHz] for $l=1,2,3,4$
and $\alpha=12.0$.
The red, blue, and purple lines represent
$T=0.001$, $0.1$, and $0.2$ [K], respectively.
The temporal periods are
$T_{0}(1)=3.02\times 10^{-6}$,
$T_{0}(2)=1.26\times 10^{-7}$,
$T_{0}(3)=6.98\times 10^{-9}$,
and
$T_{0}(4)=4.36\times 10^{-10}$ [s].
}
\label{figure09}
\end{figure}

The behavior of $F^{(-)}(t)$ for values of $l$ other than 2, namely $l=1,3,4$, as well as for cases where the influence of thermal noise cannot be neglected, is shown in Figs.~\ref{figure09}(a)-(d).
In these figures, the parameters $\omega$ and $\omega_{0}$ are chosen such that $\Delta=0$, corresponding to the resonant case.
In Fig.~\ref{figure09}(b), for $l=2$,
$T=0.001$ [K], and $t=T_{0}(2)/2$, the fidelity $F^{(-)}$ almost reaches unity.
In contrast, in Figs.~\ref{figure09}(a), (c), and (d), for $l=1,3,4$, even at the low temperature $T=0.001$ [K], the fidelity $F^{(-)}$ does not reach unity.
Furthermore, as the temperature increases, $F^{(-)}$ decreases, which is a common trend for all cases $l=1,2,3,4$.

\begin{figure}
\begin{center}
\includegraphics[width=1.0\linewidth]{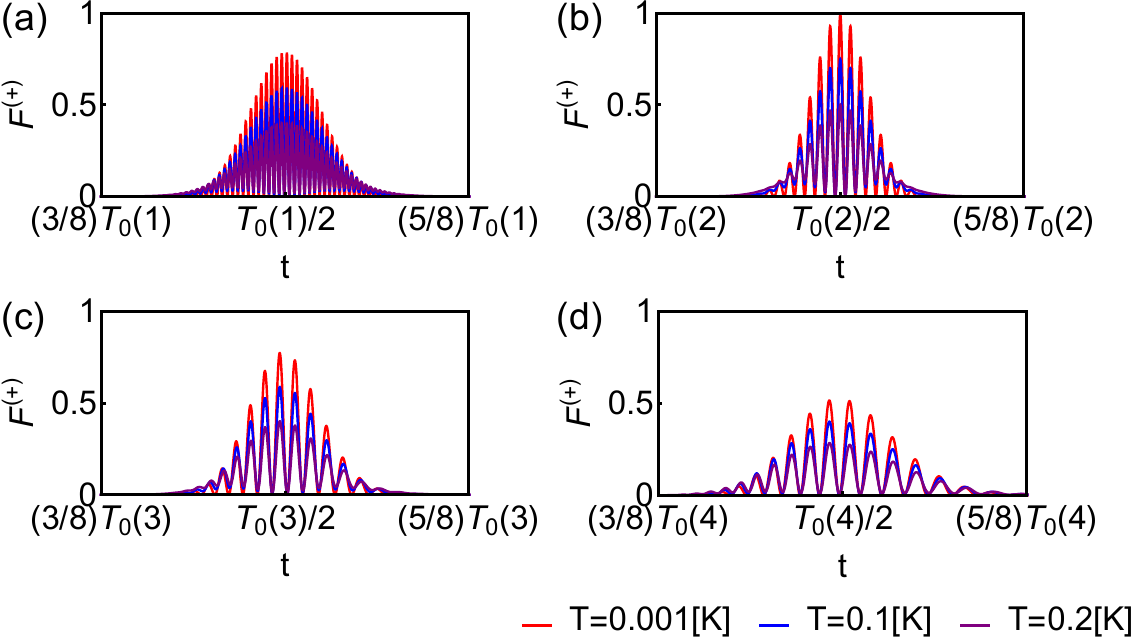}
\end{center}
\caption{
Plot of $F^{(+)}(t)$ as a function of the time variable $t$.
All physical parameters are the same as those used in Fig.~\ref{figure09}.
}
\label{figure10}
\end{figure}

The time dependence of $F^{(+)}(t)$ is shown in Fig.~\ref{figure10}.
For $l=2$, $F^{(+)}$ becomes nearly equal to unity at $t=T_{0}(2)/2$, indicating that the cat state is obtained with probability of about $1/2$.

The discussion presented thus far has assumed that $|\alpha|^{2}\gg 1$.
That is, the mechanism by which the cat state in Eq.~(\ref{2-mode-JCM-cat-state-obtained-1}) is generated relies on the assumption $|\alpha|^{2}\gg 1$.
Here, we therefore examine the dependence on $|\alpha|$ of the fidelity of the cat states approximately generated by the two-photon JCM.
Figure~\ref{figure09}(b) and Fig.~\ref{figure10}(b) show that the cat states $|\mbox{cat}^{(\pm)},\alpha\rangle$ defined in Eq.~(\ref{cat-states-definition-0}) are generated with the highest fidelity $F^{(\pm)}$ at $T_{0}(2)/2$.

\begin{figure}
\begin{center}
\includegraphics[width=0.8\linewidth]{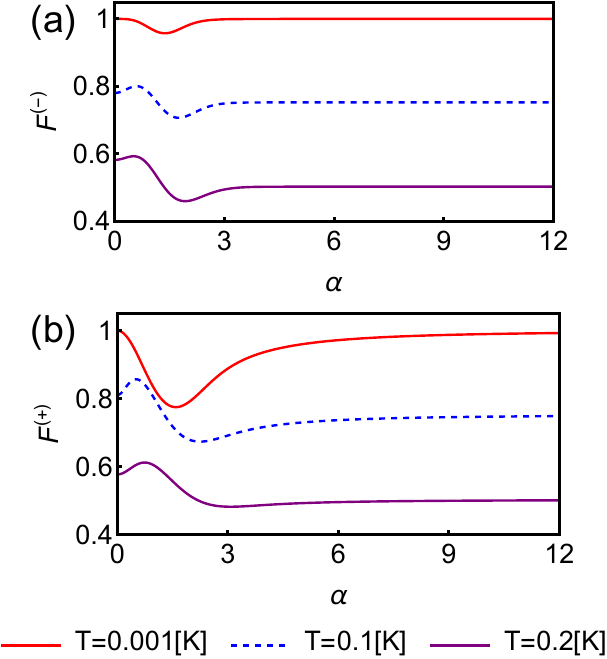}
\end{center}
\caption{
(a)
Plot of $F^{(-)}(T_{0}(2)/2)$ as a function of $\alpha$.
(b)
Plot of $F^{(+)}(T_{0}(2)/2)$ as a function of $\alpha$.
The physical parameters are set at $l=2$, $\omega/(2\pi)=2.91$ [GHz],
$\omega_{0}=2\omega$, $\Delta=0$, $g_{2}/(2\pi)=3.98$ [MHz].
The red solid, blue dashed, and purple solid lines represent $T=0.001$, $0.1$,
and $0.2$ [K], respectively.
}
\label{figure11}
\end{figure}

Figures~\ref{figure11}(a) and (b) show $F^{(-)}$ and $F^{(+)}$ at $t=T_{0}(2)/2$, respectively, plotted as functions of $\alpha$.
As can be seen from the figures, $F^{(\pm)}$ become sufficiently close to $1$ for $\alpha>7$.
Furthermore, $F^{(\pm)}$ decrease monotonically as the temperature increases.

\section{\label{section-conclusion}Conclusion}

In this paper, we have examined the concrete experimental feasibility of generating cat states by implementing the two-photon JCM in a Josephson-junction circuit.
In particular, we have numerically investigated the effects of three typical types of noise that are expected to present obstacles in experiments: magnetic flux noise with $1/f$ fluctuations, energy relaxation associated with the $T_{1}$ decay of the flux qubit, and photon loss in the bosonic mode obtained by quantizing the phase variable of the dc SQUID current.
The results show that
dissipation due to photon loss has significant effects on the fidelity of cat-state generation using the two-photon JCM.
The photon-loss calculations show how the conditional
cat-state fidelity depends on the assumed loss rate
during preparation.
In other words, we performed the photon-loss calculations assuming that the measurement of the flux qubit to obtain cat states is instantaneous and ideal.
They do not establish the quality factor $Q$ required
for a complete preparation and readout sequence
in the proposed circuit.
Cat states are among the best-known non-Gaussian states and are regarded as key components of continuous-variable quantum computers.
From this perspective, the results presented in this paper are of considerable importance.

\section*{Acknowledgment}
This paper is based on results obtained from a project,
JPNP25014,
commissioned by the New Energy and Industrial Technology
Development Organization (NEDO).

\end{document}